\documentclass[twocolumn,3p,times,procedia]{elsarticle}

\UseRawInputEncoding
\usepackage[dvipsnames]{xcolor}
\usepackage{ecrc}
\usepackage{graphicx}
\usepackage{tcolorbox}
\usepackage{titlesec}
\usepackage{soul}
\usepackage{url}
\usepackage{array,threeparttable}
\usepackage{amsmath}

\usepackage{ulem}
\volume{00}

\firstpage{1}

\runauth{}

\jid{procs}

\CopyrightLine{2022}{Published by Elsevier Ltd.}

\usepackage{amssymb}
\usepackage{amsmath}
\usepackage{multicol}
\usepackage{booktabs}

\usepackage[figuresright]{rotating}
\usepackage{bm}
\usepackage{caption}
\usepackage{fancyhdr}

\fancyheadoffset[RO,EL]{0pt}
\usepackage{geometry}

\begin{document}

\begin{frontmatter}

%% Title, authors and addresses

%% use the tnoteref command within \title for footnotes;
%% use the tnotetext command for the associated footnote;
%% use the fnref command within \author or \address for footnotes;
%% use the fntext command for the associated footnote;
%% use the corref command within \author for corresponding author footnotes;
%% use the cortext command for the associated footnote;
%% use the ead command for the email address,
%% and the form \ead[url] for the home page:
%%
%% \title{Title\tnoteref{label1}}
%% \tnotetext[label1]{}
%% \author{Name\corref{cor1}\fnref{label2}}

%% \fntext[label2]{}
%% \cortext[cor1]{}
%% \address{Address\fnref{label3}}

\dochead{}
%% Use \dochead if there is an article header, e.g. \dochead{Short communication}
%% \dochead can also be used to include a conference title, if directed by the editors
%% e.g. \dochead{17th International Conference on Dynamical Processes in Excited States of Solids}
\title{
\begin{flushleft}
{\bf  Overview of RIS-Enabled Secure Transmission in 6G Wireless Networks}
\end{flushleft}
}
%% use optional labels to link authors explicitly to addresses:
%% \author[label1,label2]{<author name>}
 %
%% \address[label2]{<address>}

\author[1]{JungSook Bae\corref{cor1}}% 
% \fnref{fn1}}
\ead{jsbae@etri.re.kr}
\author[2]{Waqas Khalid\corref{cor1}}%
% \fnref{fn2}} 
\ead{waqas283@korea.ac.kr}
\author[1]{Anseok Lee}%\fnref{fn1}} 
\ead{alee@etri.re.kr}
\author[1]{Heesoo Lee}%\fnref{fn1}} 
\ead{heelee@etri.re.kr}
\author[3]{Song Noh\corref{cor2}}%
%   \fnref{fn3}} 
\ead{songnoh@inu.ac.kr}
\author[4]{Heejung Yu\corref{cor2}}%
%   \fnref{fn4}} 
\ead{heejungyu@korea.ac.kr}

\cortext[cor1]{Co-first authors with equal contribution.}
\cortext[cor2]{Corresponding authors.}

\address[1]{Terrestrial \& Non-Terrestrial Integrated Telecommunications Research Laboratory, Electronics and Telecommunications Research Institute, Daejeon 34129, South Korea}

\address[2]{Institute of Industrial Technology, Korea University, Sejong 30019, South Korea}

\address[3]{Department of Information and Telecommunication Engineering, Incheon National University, Incheon 22012, South Korea}

\address[4]{Department of Electronics and Information Engineering, Korea University, Sejong 30019, South Korea}

%\author[]{\bf  {JungSook Bae$^{a,*}$, Khalid Waqas$^{b,*}$, Anseok Lee$^a$, Heesoo Lee$^a$, Song Noh$^{c,**}$, Heejung Yu$^{d,**}$}}

%\address{\bf \leftline {$^a$Terrestrial \& Non-Terrestrial Integrated Telecommunications Research Laboratory, Electronics and Telecommunications Research Institute, Daejeon, 34129, South Korea}

%\bf  \leftline {$^b$Institute of Industrial Technology, Korea University, Seong, 30019, South Korea}

%\bf  \leftline {$^c$Department of Information and Telecommunication Engineering, Incheon National University, Incheon 22012, South Korea}

%\bf  \leftline {$^d$Department of Electronics and Information Engineering, Korea University, Sejong, 30019, South Korea}
%}

%\cortext[]{J. Bae and W. Khalid are co-first authors with equal contribution.}
%\cortext[]{ H. Yu and S. Noh are corresponding authors. (emails:heejungyu@korea.ac.kr, songnoh@inu.ac.kr)}

%\fntext[]{J. Base, A. Lee, and H. Lee are with the Terrestrial \& Non-Terrestrial Integrated Telecommunications Research Laboratory, Electronics and Telecommunications Research Institute, Daejeon, Korea. (emails:alee@etri.re.kr, heelee@etri.re.kr).}

%\fntext[]{W. Khalid is with the Institute of Industrial Technology, Korea University, Sejong, Korea. (email:waqas283@korea.ac.kr).}

%\fntext[]{S. Noh is with the Depart. of Information and Telecommunication Engineering, Incheon National University, Incheon, Korea.}

%\fntext[]{H. Yu with the Depart. of Electronics and Information Engineering, Korea University, Sejong, Korea.}

\begin{abstract}
{As sixth-generation (6G) wireless communication networks evolve, privacy concerns are expected due to the transmission of vast amounts of security-sensitive private information. In this context, a reconfigurable intelligent surface (RIS) emerges as a promising technology capable of enhancing transmission efficiency and strengthening information security. This study demonstrates how RISs can play a crucial role in making 6G networks more secure against eavesdropping attacks. We discuss the fundamentals, and standardization aspects of RISs, along with an in-depth analysis of physical-layer security (PLS). Our discussion centers on PLS {design} using RIS, highlighting aspects like beamforming, resource allocation, artificial noise, and cooperative communications. We also identify the research issues, propose potential solutions, and explore future perspectives}. Finally, numerical results are provided to support our discussions and demonstrate the enhanced security enabled by RIS.
\end{abstract}

\begin{keyword}
6G \sep Physical-layer security (PLS) \sep Reconfigurable intelligent surface (RIS)

%% keywords here, in the form: keyword \sep keyword
%Provide 3 to 8 pieces of words or phrases to serve as guidelines for indexing \sep Using American spelling \sep  avoiding general and plural terms and multiple concepts
%% PACS codes here, in the form: \PACS code \sep code

%% or \MSC[2008] code \sep code (2000 is the default)

\end{keyword}

\end{frontmatter}

%%
%% Start line numbering here if you want
%%
% \linenumbers

%% main text

\section{Introduction}
{Sixth-generation (6G) wireless communication networks will offer significantly higher {transmission rate}, reduced latency, and enhanced reliability. These enhancements will facilitate innovative applications, unprecedented services, and comprehensive solutions \cite{AI6G_ICTE2022}. Nevertheless, 6G networks will also introduce substantial security challenges attributed to the inherent broadcast nature of wireless channels, the voluminous influx of sensitive and confidential data (e.g., personal tracking, financial transactions, and cell phone information), and the exponential surge in potential attack vectors. Therefore, 6G networks must be hyper-secure, encompassing data security from the application to the physical layer. A physical-layer security (PLS) approach based on information-theoretic principles has generated interest in industry and academia. Notably, PLS ensures data security by preventing unauthorized interception and malicious use of private information, all without encryption methods \cite{STAR_RIS_NETa33}.}

In 6G networks, sub-terahertz (sub-THz) bands will complement millimeter-wave (mmWave) bands that have been adopted in fifth-generation (5G) networks. Though high-frequency bands offer several advantages, they also pose unique propagation challenges that must be addressed. Notably, mmWave and sub-THz signals are susceptible to blockages and substantial penetration loss \cite{6G_VTM20}. Therefore, algorithms and protocols for wireless transmission must be developed to mitigate the adverse effects of an uncontrolled radio environment. Conventional transmission strategies use multiple antennas, complex signal-processing algorithms, and advanced encoding-decoding procedures. However, the transceiver design cannot be overly complex in resource-constrained scenarios \cite{STAR_RIS_rgd}. The true potential of 6G networks can be realized by deploying a radio environment that can be manipulated to optimize overall network performance. In this regard, seamless wireless connectivity in 6G networks requires novel physical-layer technologies, such as reconfigurable intelligent surfaces (RISs), relays, network-controlled repeaters (NCRs), and massive multiple-input multiple-output (mMIMO). Particularly, RIS is an emerging technology that provides a controllable and programmable radio environment \cite{RIS_COMMAG21}. Given the dependence of the PLS on environments rich in scattering, dynamically controlled channels enabled by RIS can significantly enhance security.

{
\subsection{Motivation}
A key aspect of PLS is the manipulation of the dynamic characteristics of wireless channels to enhance and/or limit the signal-to-interference-plus-noise ratio (SINR) for legitimate users and/or eavesdroppers, respectively. Due to the dependency on wireless channels characterized by noise and fading, the effectiveness of PLS might diminish under challenging propagation conditions \cite{sdt77}. Considering this concern, the dynamic channel control offered by RIS provides an opportunity to fully harness the advantages of PLS pertaining to channel propagation, spatial diversity, beamforming, and cooperative communications. By enhancing the SINR at the legitimate receiver (e.g., by diminishing fading and stabilizing the channel) and/or degrading the eavesdropping link (e.g., by inducing additional signal attenuation), RIS hinders eavesdroppers from accessing the intended message. 
Due to its significant performance benefits and alignment with the conventional PLS methods, RIS is a promising candidate to enhance PLS \cite{sdt77a}. RIS-PLS strategies find applicability within both the 5G and 6G domains. Nonetheless, safeguarding the receiver from eavesdropping attacks via RIS requires careful consideration of the strategic positioning of RIS, the optimal number of elements, and their precise configuration. Furthermore, the choice of RIS-enhanced PLS solutions depends on the specific design scenarios and communication objectives to balance security effectiveness and implementation complexity. In this study, therefore, we introduce various RIS-enabled PLS technologies and provide research issues and potential solutions.}

{
\subsection{{Summary of Goals}}
Based on the motivation above, this study demonstrates how RISs can play a crucial role in making 6G networks more secure against eavesdropping attacks. Specifically, the {goals} of the study can be summarized as follows:
\begin{itemize}
%\item We discuss the overview and standardization aspects of RISs. An overview of PLS and a description of traditional security designs are provided. 
\item We explore {RIS-enabled PLS design}, focusing on beamforming, resource allocation, antenna/node selection, artificial noise (AN), and cooperative relaying and jamming communications. 
\item We identify the research issues and potential solutions in RIS-enabled PLS. In detail, channel estimation, beam configuration, resource management, strategic placement and passive information transfer for RIS, hardware/channel modeling, and optimization for RIS-enabled PLS are discussed. 
\item In addition, we outline future research directions, highlighting machine learning (ML)-based solutions, advances in RIS hardware (e.g., active RIS, and simultaneous transmitting and reflecting-RIS (STAR-RIS)), and malicious RIS.
\item Finally, numerical results are provided to support our discussion and demonstrate the enhanced security provided by RIS. {In particular, we examine the impact of RIS modeling, the number of RIS elements, RIS beam design strategy, placement of RIS, quantized RIS phases, and AN on secrecy performance.}

\end{itemize}
}

{
\subsection{Organization}
The remainder of this paper is organized as follows: Sections 2 and 3 provide foundational discussions on RIS and PLS, respectively. Section 4 discusses {RIS-enabled PLS design}. In section 5, research issues and solutions are presented. In section 6, future directions are provided. Section 7 presents the simulation results to demonstrate the effectiveness of RIS in terms of PLS. Finally, section 8 presents the conclusions of this study.}

\begin{figure}
\centering
\includegraphics[width=.9\linewidth]{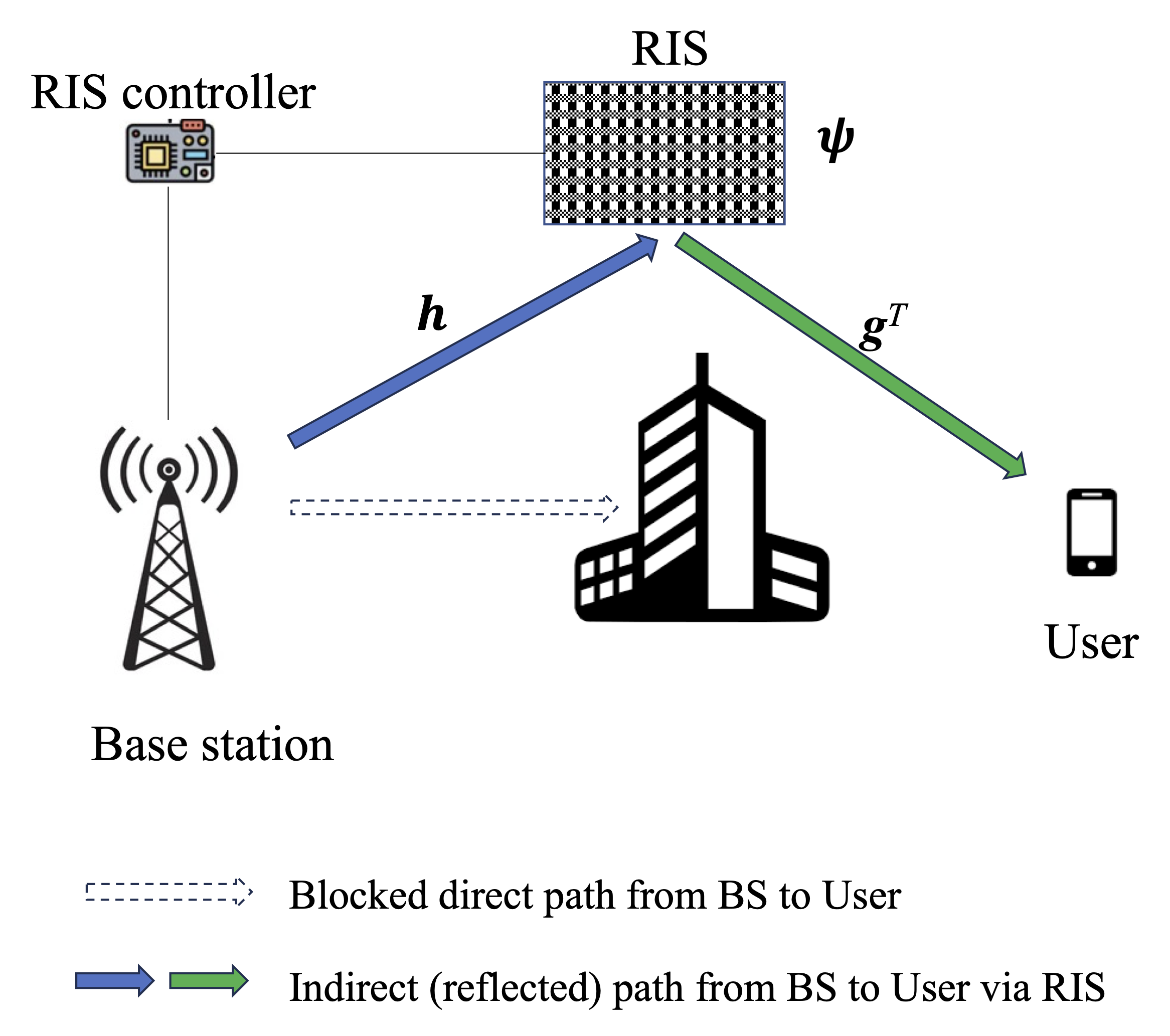}
\caption{{An RIS-based downlink transmission wireless system.}}\label{fig1}
\end{figure}

{
\section{{Reconfigurable Intelligent Surface (RIS)}}\label{secti2}

RISs have emerged as a promising hardware-based transmission technology to manipulate wireless propagation landscapes artificially. Incident electromagnetic (EM) waves are reflected by numerous flexible and discrete elements, each of sub-wavelength dimensions, embedded on a planar surface. {By tuning signal phases and/or amplitudes with a smart RIS controller, the RIS enhances the degrees of freedom (DoF) of wireless channels, improves signal transmission, and enables advanced wireless functionalities. For example, RISs can overcome unfavorable propagation conditions, such as blockage and deep fading \cite{RIS_NOMA_WCMAG20} as shown in Fig. \ref{fig1}.} A novel approach to altering radio-wave propagation, coupled with state-of-the-art signal processing at the transceiver, provides significant enhancements to wireless connectivity, including signal enhancement, interference suppression, reliable reception, and precise positioning \cite{RIS_NOMA_WCMAG20,STAR_RIS_NETaa}.}

%{RISs also offer convenient deployment (including on building surfaces and mobile/aerial vehicles), real-time configuration, full-duplex and full-band transmission, superior power-gain performance, absence of noise amplification, cost-effectiveness, and energy- and spectral-efficient attributes. On the other hand, the challenges associated with the development and implementation of RIS-aided wireless communication systems include low-complex channel estimation, hardware architecture, high-precision passive beamforming, strategic deployment, and limited control message overhead. Furthermore, several challenges are associated with the design and optimization of RIS-aided wireless communications. These challenges include hardware limitations (e.g., phase-shift quantization levels and reflecting elements), simplified assumptions in the system design (e.g., oversimplified hardware and channel models), and optimization complexities owing to highly coupled variables \cite{RIS_LOC_VTM20}.}

\subsection{{Working Principle of RIS}}

{From a hardware perspective, RISs can be constructed using metamaterials and patch arrays. RISs can be designed to serve as reflective or refractive surfaces, strategically placed to enhance communications between a base station (BS) and users \cite{R3aa}. RISs can also be classified based on energy consumption into passive-lossy, passive-lossless, or active types. For operational analysis, reflected and refracted EM waves can be characterized using equivalent models of surface electric and magnetic currents. The interaction of EM waves with RISs can be analyzed using ray-optics or wave-optics methodologies. Despite being based on approximations, they provide valuable insights into how radio waves interact with materials and are widely used in the study of RISs \cite{R4aa}.}

{For illustration, the reflection pattern of a patch-array RIS with $N$ reflection elements can be expressed by a vector $\boldsymbol{\psi}$ of which $i$th element is given by}
{\begin{equation}
\psi_i=\beta_i e^{j\theta_i}
\end{equation}
where $\beta_i$ and $\theta_i$ denote the amplitude and phase responses, respectively. As shown in Fig. \ref{fig1} where a single-antenna BS transmits a downlink signal to a single-antenna user through a RIS composed of $N$ elements, the RIS can provide an indirect (reflected) path from the BS to the user. The effective channel gain of the reflected path via the RIS is expressed by
\begin{equation}
\mathbf{g}^T \mathbf{D}_{\boldsymbol{\psi}} \mathbf{h} = \sum_{i}^{N} h_i \psi_i g_i
\end{equation}
where $\mathbf{g}^T$ and $\mathbf{h}$ denote the channels of the RIS-user and BS-RIS links, respectively, and $\mathbf{D}_{\boldsymbol{\psi}}$ is a diagonal matrix with the elements of $\boldsymbol{\psi}$ on the main diagonal. $h_i$ and $g_i$ denote the $i$th element of $\mathbf{h}$ and $\mathbf{g}$, respectively. Here, $(\cdot)^T$ denotes a transpose operation.}

\subsection{{Standardization of RIS}}
\begin{figure}
\centering
\includegraphics[width=1.0\linewidth]{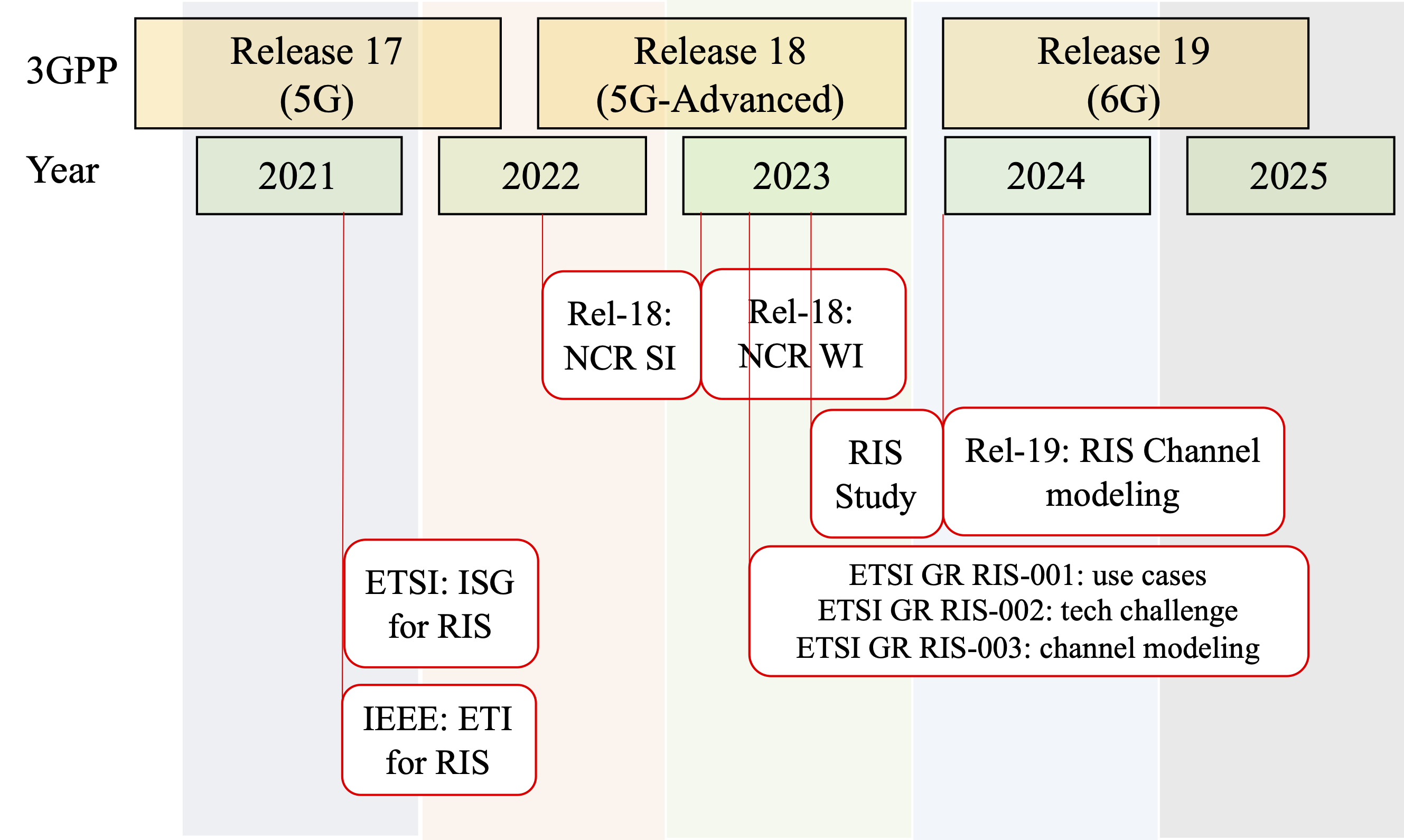}
\caption{{Standardization and related timeline for RIS.}}\label{fig2}
\end{figure}

In the International Telecommunication Union Radiocommunication Sector (ITU-R) IMT-2030 framework {document, released} in 2023, a RIS technology is discussed as one of the key enabling technologies for 6G wireless communications \cite{ITU}. In 2021, an industry specification group (ISG) for RIS was formed in ETSI {to study and standardize RIS}. {In 2023, the ISG provided three technical reports \cite{etsi_RIS1, etsi_RIS2, etsi_RIS3} as shown in Fig. \ref{fig2}.} In the Third Generation Partnership Project (3GPP) for Release 18, several companies proposed a study item for RIS. However, the majority of companies in 3GPP decided that it was too early to include RIS as a study item or working item because RIS is generally considered a candidate technology for 6G rather than 5G-Advanced. Then, it is expected to kick off a study item (SI) of RIS in 3GPP Release 19. {The progression from NCR in Release 18 to RIS in Release 19 is analyzed through comparative studies in terms of architecture, operation, control signals, etc \cite{NCR,NCR_RIS,Pot_stand}}. For technical support of the standardization of RIS, an emerging technology initiative (ETI) on RIS was formed in IEEE \cite{IEEE}. In the industry including NTT DOCOMO, preliminary field trials that demonstrate the potential of RIS in realistic environments have already been carried out \cite{NTT}. The timeline of standardization events related to RIS is illustrated in Fig. \ref{fig2}.  

%in the industry, com⁃
%panies have submitted the proposals of studying RIS in the re⁃
%lease time frame at 3GPP spanning January 2022 – June
%2023. In June 2021, an industry specification group (ISG) for
%RIS was formed in ETSI to carry out the engineering related
%study and standardization of RIS. I

%The study item may last
%over the entire duration of Release 19 (expected to end in
%December 20
%RIS has been selected as one of the potential technologies (i.e., study and working items) for 5G-Advanced and 6G cellular networks \cite{3gpp_RIS1}.  Research institutes and businesses have significantly invested in the rapid development, implementation, and evaluation of RIS prototypes. Fig. \ref{fig2} illustrates the efforts toward RIS prototyping across diverse wireless stakeholders. Such assessments have endowed researchers and engineers with direct insights into the operational efficacy of RISs in 5G/6G networks \cite{sdt77b}.

\section{Physical Layer Security}
\label{secti3}
In this section, we examine PLS, highlighting its foundational principles, traditional security designs, performance metrics, and optimization methods for secure 6G transmissions.

 \begin{figure*}[hbt!]
\centering
\includegraphics[width=\linewidth]{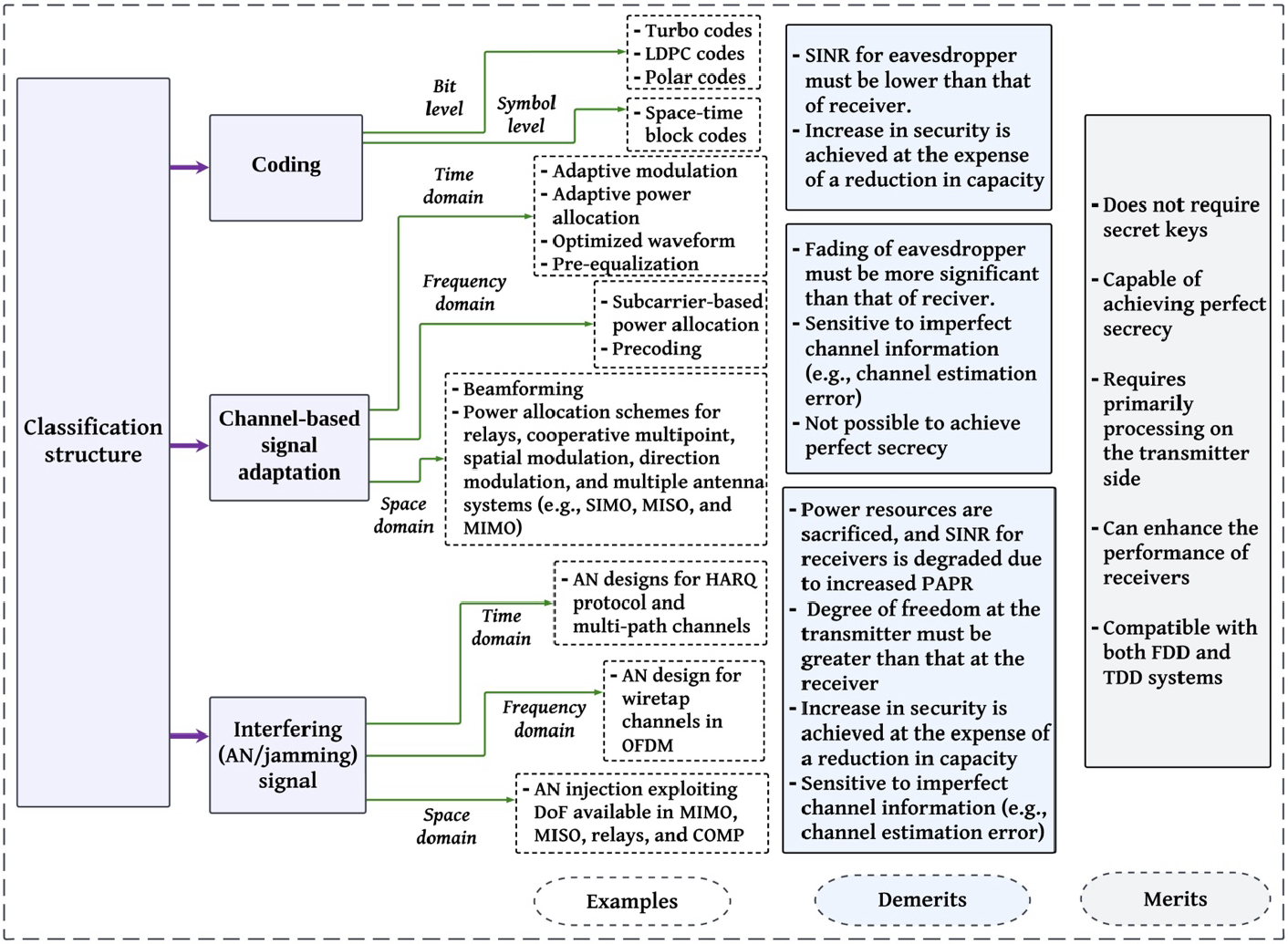}
\caption{{Classification structures (including examples, merits, and demerits) for PLS solutions in the 6G wireless environment.}}\label{fig3}
\end{figure*}

\subsection{Fundamental Concept of PLS}
Wyner investigated the concept of perfect secrecy for a wiretap channel at the physical layer by exploiting the capacity difference between legitimate and eavesdropping channels. With the rapid development of coding theories and the practical limitations of encryption-based security technologies, PLS has emerged as a leading method of enhancing secure communication \cite{5G_SECURITY_VTM20}. Unlike cryptographic methods, PLS techniques do not require the design, storage, or distribution of keys, resulting in a more cost-effective electronic security method. PLS schemes are intended to prevent eavesdropping by taking advantage of channel fading (despite its negative impact on reliability). By leveraging physical-layer characteristics, PLS methods refine transmission strategies and parameters, adapting to challenging channel conditions \cite{sdsd3}. Additionally, PLS is distinguished by its reduced complexity and resource efficiency. PLS categorizes eavesdropping scenarios into passive and active. Passive eavesdroppers passively intercept (decode/analyze) transmissions from legitimate users without initiating active actions, such as transmitting signals. Conversely, active eavesdroppers engage in both interception and adversarial activities, such as generating AN or jamming attacks, disseminating deceptive information feedback, and exploiting pilot contamination. In Fig. \ref{fig3}, a classification structure for PLS solutions tailored to 6G wireless communications is illustrated, as well as representative examples and their merits and demerits. Each security strategy has its own unique strengths and limitations, making it particularly suitable for certain applications, systems, scenarios, and channel conditions. Therefore, a combination of these approaches is expected to provide superior security enhancements compared to an individual approach.

\subsection{Performance Metrics And Optimization Methods}
From an information-theoretic viewpoint, PLS is often characterized by diverse secrecy performance metrics and objective functions \cite{5G_SECURITY_VTM20}. Table \ref{table1} provides a comprehensive breakdown of those employed for evaluating secure transmission. Furthermore, secure systems have been examined in terms of signal processing and optimization \cite{sdsd3}. In the context of PLS, a variety of wiretap channel models have been described \cite{PLS_V2X_VTM20}, as detailed below:
\begin{itemize}
\item MIMO wiretap channels: Encompassing a transmitter, a receiver, and an eavesdropper, all equipped with multiple antennas. 
\item Broadcast wiretap channels: Characterized by a single transmitter, multiple receivers, and several eavesdroppers. 
\item Relay wiretap channels: A cooperative framework featuring a transmitter, a relay, a receiver, and an eavesdropper. 
\item Multiple-access wiretap channels: Networks that include multiple transmitters, a single receiver, and an eavesdropper. 
\item Interference wiretap channels: Networks typified by multiple active communication links.
\end{itemize}

\begin{table*}[t!]
\caption{ Performance metrics for evaluating secure transmission in 6G networks.}
\footnotesize
\centering
%% \tablesize{} %% You can specify the fontsize here, e.g., \tablesize{\footnotesize}. If commented out \small will be used.

\begin{tabular}{ |p{2.6cm}| p{8cm}|  p{3.5cm}| }
\toprule
\textbf{{Performance metric}}& \textbf{{Description}} & \textbf{{Mathematical representation}} \\
\midrule
Secrecy rate
&
\begin{itemize}
\item Defined as the difference between the achievable rate of a legitimate link $\left(C_l\right)$ and that of an eavesdropping link $\left(C_e\right)$. 
\end{itemize}
&
$C_s=\max \left(C_l-C_e, 0\right)$ \\ 
\midrule
Secrecy outage probability
&
\begin{itemize}
\item Probability that the secrecy rate $\left(C_s\right)$ falls below a predefined target $\left(C_{target}\right)$.
\end{itemize}
&
$P_{out}=\Pr\left(C_s < C_{target}\right)$ \\ 
\midrule
Intercept probability
&
\begin{itemize}
\item 	It is defined with a target secrecy rate of zero offering a worst-case evaluation of system security.
\item 	It quantifies the likelihood of observing a negative secrecy rate, implying that $C_s$ consistently falls below zero. 
\end{itemize}
&
$P_{intercept}=\Pr\left(C_s \leq 0\right)$  \\ 
\midrule
Strictly positive secrecy capacity
&
\begin{itemize}
\item 	Another special case of $P_{out}$ and is defined as the probability that a non-zero $C_s$ exists. 
\end{itemize}
&
$P_{positive}=\Pr\left(C_s > 0\right)$  \\ 
\midrule
Secrecy coverage probability
&
\begin{itemize}
\item 	It measures the success of the secure delivery process and has an opposite definition to the $P_{out}$.
\item 	It is defined as the probability that $C_s$ exceeds the $C_{target}$.
\end{itemize}
&
$P_{coverage}=\Pr\left(C_s > C_ {target}\right) = 1-P_{out}$\\
\midrule
Secure energy efficiency
&
\begin{itemize}
\item Number of secured bits transferred $\left(B\right)$ per unit of energy or the total energy $\left(E\right)$ required for sending a bit with secrecy. 
\end{itemize}
&
$SEE=\frac{B}{E}$\\
\midrule
Secure power consumption
&
\begin{itemize}
\item Minimum amount of power required to achieve a specified $C_{target}$ and $E_{secure}$ is the corresponding energy consumed over the transmission time $\left(t\right)$.
\end{itemize}
&
 $P_{secure}=\frac{E_{secure}}{t}$ subject to $ C_s \geq C_{target}$ \\
\bottomrule
\end{tabular}
\label{table1}
\end{table*}
\normalsize

Based on the aforementioned secrecy performance metrics, various optimization problems can be formulated \cite{sdsd3, PLS_V2X_VTM20}. For a typical example, we can consider the maximization of secrecy performance by controlling resources, such as transmit power, beamformers (or precoders), transmission duration, and bandwidth allocation. Heuristic algorithms, such as randomized algorithms, can solve non-convex optimization problems. A non-convex or intractable problem can be transformed into a tractable convex problem by approximation and relaxation methods and can be solved by convex optimization algorithms, such as interior point methods. Moreover, quadratic programming, mixed-integer programming (to solve problems with discrete and continuous variables), alternative optimization (an iterative method for solving convex sub-problems), fractional programming (to solve the ratio of nonlinear functions), and semidefinite programming have been employed to optimize PLS problems \cite{Sec_5G_COMST20}. Furthermore, deep learning (DL) approaches can be adopted to address intricate network challenges, such as multi-cell and multi-user scenarios.

\section{{RIS-Enabled PLS Design}}
\label{secti5}

The risk of wiretapping is inherent in the vast and diverse ecosystem of the 6G networks. Traditional PLS techniques (e.g., mMIMO and cooperative communication) are often challenged by unpredictable propagation conditions. Under such scenarios, RIS-aided PLS approaches outperform those without RIS integration \cite{Sec_RIS_COML19}. Such scenarios could arise when the receiver requires a high level of secrecy, eavesdroppers are more numerous, equipped with superior antenna setups, control a dominant channel, exhibit a strong correlation with the receiver, or are located closer to the transmitter than the receiver \cite{Sec_5G_COMST20, Sec_RIS_COML19}. In such circumstances, it may not be possible to exploit spatial DoF for secrecy enhancement through transmit beamforming with large-scale antenna arrays. In addition, hybrid techniques, such as transmit beamforming with AN or cooperative jamming, may not always prove effective in weakening the reception of an eavesdropper \cite{PLS_V2X_VTM20, Sec_5G_COMST20, Sec_RIS_COML19}. 

Using RIS to adjust the reflection of signals, coupled with signal processing optimization at both ends, offers distinct security benefits. Flexible signal adjustment via intelligent passive elements addresses severe fading in traditional channels. {An optimally configured RIS can enhance wireless channel efficiency for secure communications, regardless of the number, position, and channel state of transmitters, receivers, and potential eavesdroppers \cite{Sec_MIMO_COML20}. In typical RIS-aided PLS systems, a transmitter can send a message to a receiver via an RIS in the event of an eavesdropping attack. When an RIS is positioned closer to a transmitter or receiver, PLS performance can be enhanced, which is primarily influenced by the number of reflecting elements. Several studies have demonstrated that increasing the number of reflecting elements within an RIS is more effective than expanding the antenna array at a transmitter in ensuring transmission security \cite{Sec_RIS_COML19, Sec_MIMO_COML20}. The PLS performance may deteriorate due to the fact that eavesdroppers may receive multiple copies of the intended signals through the RIS, resulting in a severe leak of information. In these scenarios, however, relying solely on a secure technique may not provide adequate protection against eavesdropping. The implementation of a synergistic strategy for reducing the quality of the signal for the eavesdropper while enhancing the quality of the signal for the receiver can yield significant security improvements.}

 \begin{figure*}[hbt!]
\centering
\includegraphics[width=4.5in]{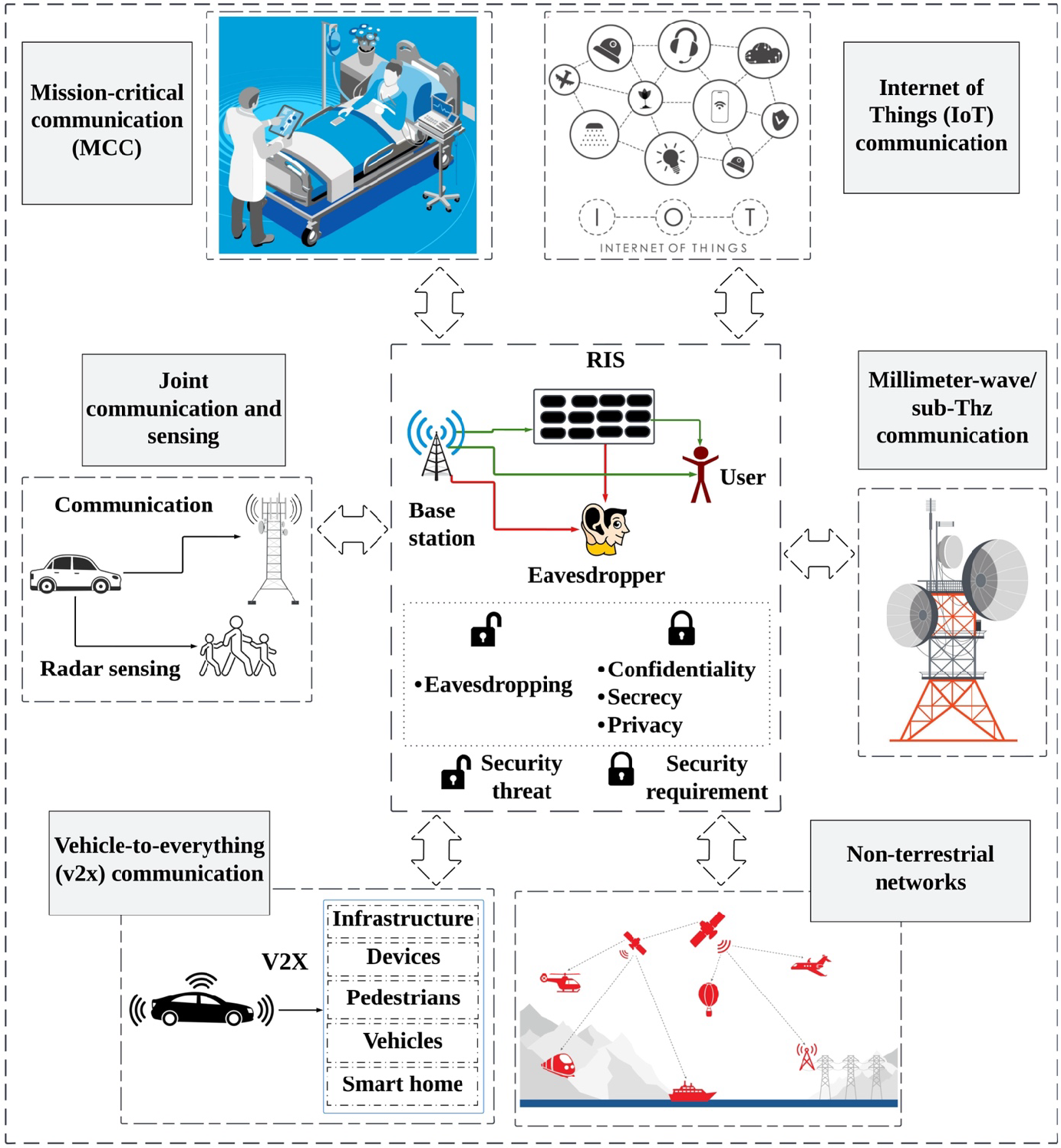}
\caption{{Illustration of the 6G wireless paradigm in which RIS-PLS can play a crucial role.}}\label{fig4}
\end{figure*}

In terms of secure transmissions, RIS systems offer compelling advantages \cite{Sec_RIS_COML19, Sec_MIMO_COML20}. As illustrated in Fig. \ref{fig4}, PLS designs based on RIS can be integrated into key 6G technologies, such as mission-critical communications (MCC), non-terrestrial networks, vehicle-to-everything (V2X), joint communication and sensing, mm-wave/sub-THz communications, and Internet of Things (IoT) communications. However, research on RIS-aided PLS solutions within 6G networks remains in its infancy \cite{Waqas_ACCESS21}.  {Further investigation is essential to understand the strengths and limitations of design solutions across diverse network topologies and application scenarios, focusing on their implementation efficacy, inherent complexity, and control variables. Based on this, we present RIS-enhanced PLS designs focusing on resource allocation, beamforming, antenna/node selection, AN generation, and cooperative methods.}

\subsection{Resource Allocation in RIS-Enabled PLS}
Secure resource allocation involves the use of network resources, including frequency bandwidth, time slots, and power levels, to ensure secure transmission \cite{Waqas_SENSOR20}. PLS provisioning can be achieved by using subcarrier allocation, adaptive power allocation, or a combined approach of subcarrier and power allocation. {By adapting the link between the transmitter and receiver via RIS by adjusting transmission parameters, a secure communication link can be established. RIS can direct signals to the receiver and/or degrade the eavesdropper's SNR. In particular, RIS-based link adaptation and channel-dependent resource allocation can be designed to provide flexible and scenario-specific secure transmission. In this context, parameters adjustment based on channel characteristics include transmit power, the number of RIS elements and reflection coefficients, subcarriers, and channel bandwidth \cite{ddd2t}.}

\subsection{Beamforming in RIS-Enabled PLS}
The spatial DoF offered by multiple antennas in MIMO systems enhances both the reliability and security of data transmission. Utilizing beamforming and precoding techniques, spatially focused signals can be strategically transmitted to realize diversity and array gains. Specifically, beamforming applies to rank-one transmissions, where a single data stream is transmitted via a multi-antenna array. Conversely, precoding involves multi-rank transmissions, signifying the concurrent transmission of multiple data streams \cite{Waqas_SENSOR21, Yu_TIFS18}. A robust security strategy entails the mathematical optimization of beamforming and precoding vectors to fulfill predefined PLS design criteria.

{The passive beamforming capabilities of the RIS can synergistically complement the active beamforming techniques employed by the transmitter to enhance PLS performance metrics. For instance, through the joint optimization of the transmit beamforming vectors and the phase shift design of the RIS elements, it is possible to strategically degrade the eavesdropping channel relative to the legitimate channel, while concurrently enhancing the decoding signal strength at the receiver. Secure transmission efficiency can be improved by increasing the number of RIS elements, rather than by enlarging the transmitter's antenna array. Hence, when complemented by RIS deployment, a reduced antenna count at the transmitter will result in significant secrecy gains \cite{ddd2t2}. Enhanced secure beamforming via RIS integration can improve PLS in various setups, such as multiple data streams, multiple users, or wide frequency bands. The optimal approach, however, is determined by the rank of the transmitted data and the level of security requirements.}

\subsection{Antenna/Node Selection in RIS-Enabled PLS}

The selection of antennas or nodes in MIMO systems plays a critical role in optimizing the system performance, including aspects such as spectral efficiency and SNR \cite{sdt3, Anti_jam_IOTJ20}. {With RIS-enabled PLS, antenna/node selection can provide a complementary advantage. Specifically, while RIS elements manipulate the wireless propagation environment to secure the channel against eavesdroppers, an optimized antenna/node selection in the MIMO system can further enhance this security paradigm by adaptively choosing the best set of antennas and nodes to transmit. Combining these strategies allows for sophisticated manipulation of channel states between receivers and eavesdroppers. Therefore, it creates a multi-layered security approach that not only meets predefined PLS design criteria but also optimizes the tradeoff between security and system performance. Thus, integrating antenna/node selection with RIS-enabled PLS is essential to achieve a comprehensive and robust secure communication system.}

\subsection{AN Generation in RIS-Enabled PLS}
AN can be generated by either the transmitter or the receiver to mitigate eavesdropping attacks \cite{Yu_TCOM21}. Specifically, the transmitter can transmit AN within the same frequency band as the legitimate signal by leveraging the null space in the channel. Alternatively, the receiver can generate AN via in-band full-duplex (FD) communications. {RIS serves as a promising countermeasure by reflecting AN, thereby intensifying interference experienced by eavesdroppers. A RIS-aided AN design can achieve {an} equivalent secrecy level with fewer elements and reduced computational complexity compared to a design without AN \cite{AN_WCL20}. However, AN methods are power-intensive, requiring a delicate balance of transmit power for both secure and reliable communications. By utilizing RIS, {power constraints} may be alleviated while maintaining higher communication performance. Optimizing RIS-based AN generally involves selecting optimal AN power using real-time channel state information (CSI) and determining the phase shifts for the RIS.}

\subsection{Cooperative Relaying and Jamming in RIS-Enabled PLS}
Spatial diversity in cooperative relaying and jamming enhances efficient security measures. In cooperative relaying, trusted relays provide the diversity benefits of MIMO, improving PLS. As outlined in Table \ref{table2}, amplify-and-forward (AF) and decode-and-forward (DF) are commonly employed relaying protocols. Cooperative jamming strategically employs relays to send disruptive signals, undermining eavesdropper interceptions \cite{Se_multihop_TIFS21}. Nevertheless, cooperative relaying raises unresolved challenges, such as relay selection, reliability, power management, positioning, and computational burden. Cooperative jamming challenges include incentive policy, power allocation under imperfect CSI, and jamming signal design against multiple eavesdroppers \cite{PLS_COMST19}.

{Distinct from cooperative relaying and jamming approaches, an RIS-enabled architecture boasts enhanced spectral and energy efficiencies, facilitated by FD operations. Moreover, RIS passive attributes obviate the need for additional phase or interference cancellation techniques, resulting in a significant advancement over conventional cooperative approaches. In the domain of PLS enhancement methodologies, RIS can also operate in an integrated manner with cooperative relaying and jamming techniques \cite{sdt6}. Specifically, RIS can tailor the propagation environment to improve the SNR at the receiver or degrade it at an eavesdropper, while cooperative relays can forward the intended signals over multiple paths, adding an extra layer of security through spatial diversity. Alternatively, cooperative jamming can disrupt eavesdropper reception selectively, without compromising legitimate communication. Through leveraging cooperative diversity benefits and RIS reconfigurability, the system can adaptively modify its transmission and jamming strategies to meet the dynamic security requirements of 6G networks.}

\begin{table}[t!]
\caption{Comparison of RIS vs. DF and AF relays.}
\footnotesize
\centering
%% \tablesize{} %% You can specify the fontsize here, e.g., \tablesize{\footnotesize}. If commented out \small will be used.

\begin{tabular}{ |p{2.2cm}| p{0.8cm}|  p{1.2cm}| p{1.6cm}|}
\toprule
\textbf{{Attributes }}& \textbf{{RIS}} & \textbf{{DF relay}} &  \textbf{{AF relay}}\\
\midrule
Hardware cost
&
Low
&
High  & Intermediate\\ 
\midrule
Duplex
&
Full
&
Half & Full/Half \\ 
\midrule
Power consumption
&
Low
&
High  & Intermediate \\ 
\midrule
Noise amplification
&
No
&
No  & Yes \\ 
\midrule
Complicated signal processing
&
No
&
Yes & No\\
\midrule
RF chain
&
No
&
Yes & Yes\\
\bottomrule
\end{tabular}
\label{table2}
\end{table}
\normalsize

\section{Research Issues and Potential Solutions in RIS-Enabled PLS}
\label{secti5}
In this section, we discuss the research issues and potential solutions related to the design and implementation of secure 6G wireless networks using RIS.

\subsection{Estimation of Channels Involving RIS}
\emph{Research Issues}: In RIS-enabled frameworks, PLS enhancement depends on the precise reconfiguration of the RIS elements, which requires accurate, timely, and low-complex channel estimation. While the theoretical upper bound of performance can be obtained with the assumption of perfect CSI for both receiver and eavesdropper, obtaining such perfect CSI presents formidable challenges \cite{YU_AN:SYSJ19}. The challenges include hardware limitations, non-linear characteristics of RIS elements, lack of information for passive eavesdroppers, and channel estimation errors \cite{YU_NOMA_AJ:FGCS20}.

\emph{Potential Solutions}: Under varying system architectures and channel conditions, channel estimation problems have been studied in RIS-enhanced secure systems \cite{CE_RIS_TWC23}. By employing training signals, low-power receiving RF chains at the RIS can be used to estimate individual channels between the transmitter and the RIS, as well as between the RIS and the receiver. Extrapolation-based methods offer enhanced accuracy in spatially sparse channels, especially in mm-wave and sub-THz bands, although the development of reliable hardware and channel modeling remains a priority. Using ML and sparsity-aware algorithms can reduce pilot overhead. For example, the transmitter-RIS channel, which has more unknown coefficients due to the greater number of antennas at the transmitter, can be estimated less frequently than the more dynamic RIS-receiver channel. In setups where RIS lacks receiving RF chains, another method is to estimate the concatenated transmitter-receiver channel via RIS, taking advantage of assumed uniform configurations among nearby reflective elements to minimize computational complexity, but at the expense of degraded estimation accuracy \cite{qaa1}. In future 6G networks, a complex network topology, dense deployment of multiple RISs, and the unique propagation characteristics of mm-wave/sub-THz bands will make real-time estimation of RIS channels challenging.

\subsection{Beam Configuration in RIS-Enabled PLS}
\emph{Research Issues}: In RIS-enabled PLS systems, beam configuration is implemented to optimize signal propagation and enhance security \cite{STdsd}. Nevertheless, it presents a number of challenges. Advanced algorithms are necessary for optimizing RIS orientation when dealing with highly directional beams. Coherent signal processing requires synchronization between RIS elements and existing transceivers. Dynamic changes in the environment require quick reconfiguration algorithms and robust channel estimation methods. Finally, beamforming efficacy is constrained by hardware limitations, such as phase quantization errors and spatial correlation \cite{sdt5}.

\emph{Potential Solutions}: For adaptive beamforming under time-varying channels, ML algorithms are emerging as viable solutions. Time and phase coordination can be developed with advanced clock synchronization methods. The use of robust optimization techniques is promising for adapting to environmental changes, and hybrid channel estimation methods can be developed to improve CSI accuracy. Overall, advancements in ML, optimization algorithms, and synchronization techniques offer increasingly effective solutions for beam configuration in RIS-enabled PLS systems.

\subsection{Resource Management in RIS-Enabled PLS}
\emph{Research Issues}: Future 6G networks may have complex scenarios due to large-scale and random deployment of RIS, a transmitter, and multiple receivers and eavesdroppers with large antenna arrays \cite{RISENV_COMMAG20}. In such scenarios, centralized transmission is not recommended due to the high feedback overhead, computational complexity, and high energy consumption.

\emph{Potential Solutions}: The development of distributed algorithms is becoming increasingly important in managing complex scenarios. These can be tailored to optimize various network functionalities, such as active beamforming at the transmitter, passive beamforming at the RIS, and relay selection/scheduling in cooperative PLS methods. Due to the massive scale and complexity of future 6G networks, the design, configuration, and operation of distributed architectures will remain a formidable challenge.

\subsection{Placement of the RIS for Enhanced PLS}
\emph{Research Issues}: Despite their shorter coverage range compared to active relays, RISs pose unique challenges in PLS within hybrid 6G networks (having both passive RISs and active transmitters). In particular, passive eavesdroppers may exploit spatial correlations or beamforming errors to intercept confidential communications. RIS placement can mitigate this issue by increasing the SNR at the receiver while minimizing the SNR at the eavesdropper \cite{sdt5, RISENV_COMMAG20}.

\emph{Potential Solutions}: Optimal RIS placement in the context of PLS can be achieved by utilizing optimization algorithms that take into account the geometrical relationship between the RIS, the transmitter, the receiver, and the eavesdropper. When full CSI is available, exhaustive search methods can be used. A heuristic approach can be used for simpler configurations (e.g., single-cell scenarios) \cite{ddd2}. ML approaches enable the deployment of RIS in complicated network topologies and the adaptive configuration of RIS elements in {real-time}, providing the network with greater protection against eavesdropping threats. Due to the complexity involved in estimating CSI involving RISs in future 6G networks, especially in scenarios involving multiple transmitters and RISs, ML approaches are becoming increasingly relevant.

 \subsection{Passive Information Transfer for the RIS}
\emph{Research Issues}: An RIS has information about its control signals (to coordinate with a transceiver), maintenance signals (to ensure correct operation), and feedback signals (to transmit the estimated CSI). To develop an RIS setup for PLS enhancement, this information must be passively transmitted.

\emph{Potential Solutions}: In \cite{PBIT_WCL20}, the authors proposed a joint passive beamforming and information transfer (PBIT) approach to transmit RIS information passively without consuming additional resources. However, there has been no investigation of the inherent tradeoff between passive information transfer and passive beamforming designs for secure transmission in PBIT. Furthermore, PBIT is characterized by stochastic optimization problems that are difficult to resolve.

 \subsection{Hardware and Channel Modeling for RIS-Enabled PLS}
\emph{Research Issues}: The development of unified and physics-compliant hardware models for RIS-enabled PLS constitutes a dynamic research frontier. It involves evaluating RIS functionality across hardware architectures and investigating its interaction with arbitrary EM fields. RIS models must rigorously consider element coupling, impedance matching, hardware imperfections, and scattering properties. Additionally, channel models serve as a bridge between hardware specifications and communication theory, providing mathematical formulations of the complex interaction between wireless signals and radio environments. For large-scale and small-scale channel characteristics, conventional channel models utilize path loss and multi-path fading, respectively. However, 6G ecosystems will require rigorously validated channel models that are specifically tailored to RIS-enabled PLS.

\emph{Potential Solutions}: Phase shift, load impedance, and generalized sheet transition condition models can characterize RIS reconfigurability and define RIS-EM field boundary conditions. Innovative hardware designs and manufacturing solutions are essential to increase the scalability and cost-effectiveness of RISs while maintaining their tunability and real-time controls \cite{HolographicMIMO_WCM20}. Both amplitude-phase and phase-only controls can use quasi-continuous quantization, with a trade-off between implementation complexity and performance. Furthermore, a unified channel modeling framework is imperative for diverse application scenarios, including mMIMO, underwater communication, and satellite networks. Path loss models must incorporate design-specific scaling variables, such as RIS size, element arrangement, and transmission distance. As 6G RIS systems move towards higher frequencies and enlarged apertures, indoor wireless, and near-field propagation will become more critical. There is a noticeable absence of empirically based, mathematically robust channel models suitable for evaluating physical-layer performance under diverse deployment scenarios, architectural paradigms, and system parameters. Generally, channel models can be categorized into deterministic and stochastic types. Mathematical approximations are also necessary for complicated fading distributions. 

 \subsection{Optimization for RIS-Enabled PLS}
\emph{Research Issues}: Due to multiple and coupled variables (related to the transmit beamforming, passive beamforming, and relays/jammers in cooperative scenarios), RIS-aided PLS designs present an intricate mathematical challenge. Joint optimization of these variables is typically intractable. Additionally, non-convex constraints, such as the source power constraint, unit-modulus constraint, and discrete phase adjustment constraint, complicate the optimization process. Security is generally improved by large-scale RIS. However, it increases the dimensions of phase-shift matrices, adding a computational burden. Multi-user and multi-element RIS configurations require a large channel space, which makes fading channel conditions difficult to describe mathematically. In addition, 6G systems will have complex topologies and nonlinear components. Thus, advanced optimization techniques are required to design algorithms that manage non-convexity with minimal signaling overhead \cite{sdt1, Wcdfasds}.

\emph{Potential Solutions}: RIS-enabled 6G systems present a diverse and complex optimization landscape, rendering traditional convex methods like linear programming or dynamic programming ineffective. To address high-dimensional or multi-objective optimization problems, robust strategies are needed. Hardware limitations and channel imperfections such as the phase-dependent amplitude of the RIS, transceiver distortions, and CSI error can be ignored for simplification. Through relaxation techniques, intractable non-convex problems can be approximated analytically. Using iterative or heuristic algorithms, non-convex objectives can be approximated more accurately into convex ones.

 \begin{figure*}[hbt!]
\centering
\includegraphics[width=5in]{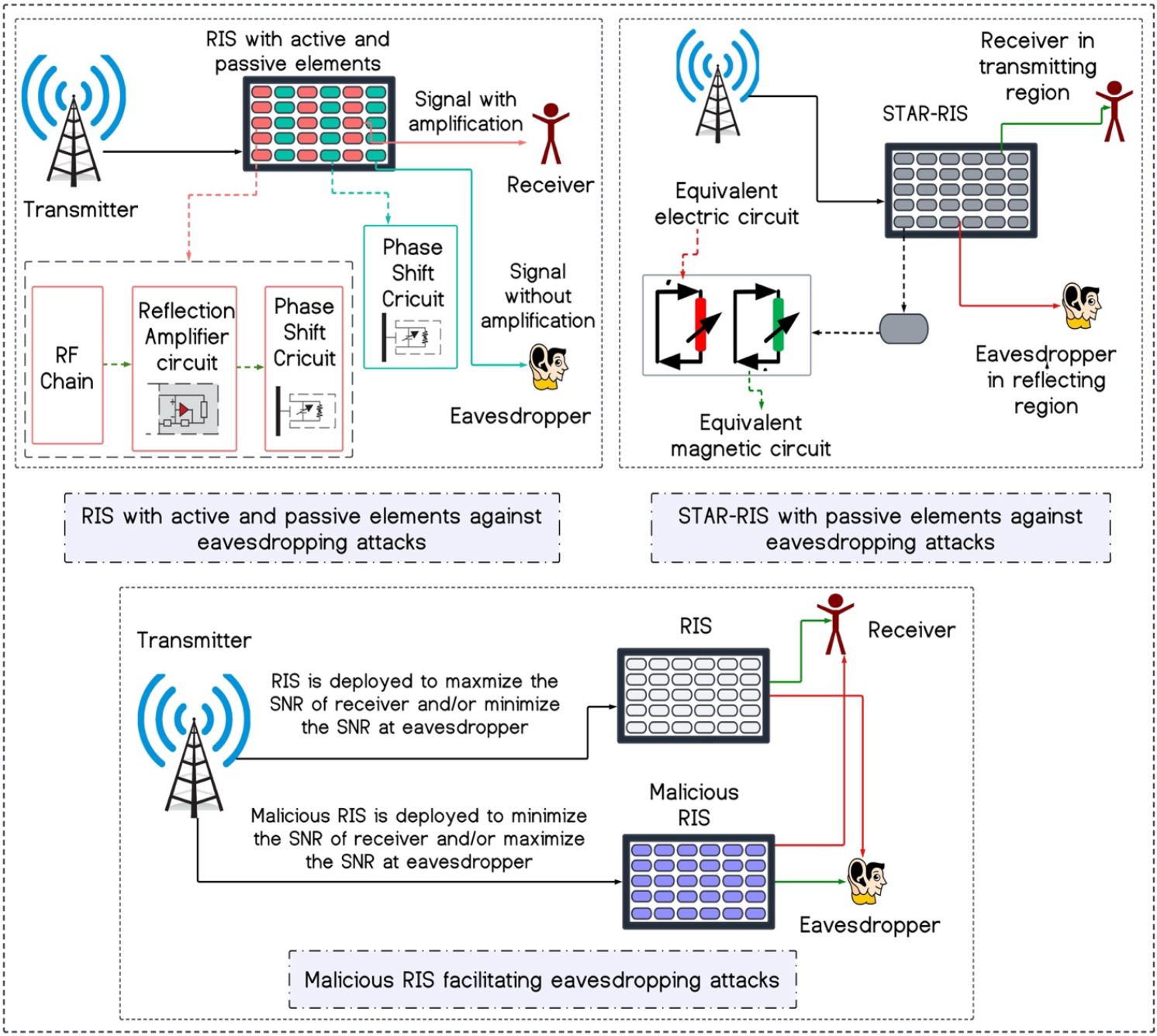}
\caption{{Active RIS, STAR-RIS, and malicious RIS in the context of RIS-enabled PLS.}}\label{fig5}
\end{figure*}

\section{Future Perspectives and Research Avenues for RIS-Enabled PLS}
This section outlines future research avenues, focusing on ML-based solutions, advanced RIS types, and malicious RISs. These topics are clarified in the context of RIS-enabled PLS as illustrated in Fig. \ref{fig5}.

 \subsection{ML-Based Solutions}
Analytical approaches in RIS-aided PLS often employ complex mathematical models that are limited in applicability and adaptability due to rigid assumptions. Conventional methods often fail in cases involving variable RIS configurations, dynamic channel states, and adversarial user behavior. RIS configurations can be optimized for security through ML algorithms, particularly DL and deep reinforcement learning (DRL) \cite{sdt13}. These algorithms learn network environments and dynamically adapt RIS settings to create secure communication links. This makes eavesdropping significantly more challenging because it introduces high dimensionality and randomness into the channel response. DL- and DRL-based approaches offer robust solutions to various aspects of RIS-PLS in 6G systems, including transmitter-side beamforming, receiver-side optimization, and RIS-side passive beamforming. With DL, channel estimation and tracking are improved regardless of imperfect conditions by leveraging model-free mappings and learnable parameters. DRL has the advantage of large search spaces and the ability to optimize multi-objective problems, such as coverage and secrecy. However, RIS-aided secure 6G systems have expansive channel spaces, which impose computational demands on DRL algorithms. Therefore, further research is necessary to create learning algorithms that reduce computational overhead, ensure real-time adaptability, and maintain accuracy and stability. In this context, the future direction may lie in a holistic approach to RIS-enabled PLS in 6G networks, necessitating the co-design of ML algorithms and cryptographic measures.

\subsection{Advances in RIS Hardware}
 
\emph{Active RIS}: To mitigate double-fading attenuation, active RIS has been proposed \cite{Wsdsd1}. Unlike passive RISs, active RISs reflect incident signals with adjustable phase shifts as well as amplify them. In active RISs, phase shift circuits and reflection-type amplifiers are embedded within the architecture to convert multiplicative channel loss into an additive form and augment it with amplification gain. It is possible to achieve optimal system performance through the combination of active and passive RIS elements, although power consumption must be considered. Active RISs have lower hardware overhead than traditional relays, which require components such as digital-to-analog converters, mixers, and power amplifiers. Under equivalent conditions, active RISs offer better performance-to-overhead ratios since they use only power amplification and diodes. While RIS-aided secure communications have been studied, the impacts of the deployment and use of active RISs on secrecy rate are unknown. Hence, future work should address these challenges and investigate the efficiency and cost-effectiveness of active RISs.

\emph{STAR-RIS}: The transmitter and receiver must be on the same side of a reflecting-only RIS system, limiting receiver coverage behind the RIS. STAR-RISs can provide additional DoF by modifying signals simultaneously in full space \cite{STAR_RIS_NETMAG23}. In \cite{BD-RIS}, STAR-RIS is categorized as a special case of beyond diagonal RIS (BD-RIS) \cite{BD-RIS}. STAR-RISs support both transmission (T) and reflection (R) functionalities. In its hardware design, STAR-RIS uses electric and magnetic currents to facilitate simultaneous or sequential T and R signals. To enable independent or coupled communication, STAR-RIS uses three operating protocols, namely energy splitting, mode selection, and time splitting. Adapting RIS-based PLS strategies to a STAR-RIS-based PLS framework remains challenging. This is primarily due to the performance analysis incorporating newly introduced tunable parameters for both T and R links, hardware tuning mechanisms for T and R elements, and comprehensive channel modeling in full space. In particular, deployment strategies for STAR-RIS and multi-user beamforming should be further investigated. It is essential to select an optimal STAR-RIS protocol that balances complexity and performance. Furthermore, intrinsic theoretical constraints, such as coupled phase and unit modulus constraints complicate PLS optimization. To address these issues, it is necessary to design ML algorithms and optimized convex algorithms that can handle hybrid control schemes incorporating both continuous and discrete variables.

 \subsection{Malicious RISs}
Legitimate and malicious RISs can coexist in the emerging 6G landscape. By maximizing information leakage, malicious RISs compromise security, increasing the eavesdropper's SNR rather than the receiver's. PLS challenges are complicated by illicit data transmission and pilot contamination. Due to malicious RIS's passive nature and lack of CSI, legitimate RISs cannot nullify their signal impact, reducing the accuracy of channel estimation for legitimate links and undermining pilot-based CSI techniques. Therefore, existing optimization strategies based on instantaneous CSI become outdated and ineffective \cite{sdt145}. In the presence of malicious RISs, comprehensive analysis, and countermeasures are needed for secure RIS-aided 6G networks. In future research, optimization frameworks can be developed to counter eavesdropping even without perfect CSI. Furthermore, empirical case studies and performance evaluations are essential to validate PLS solutions against malicious RISs.

	\begin{figure}
\centering
\includegraphics[width=1\linewidth]{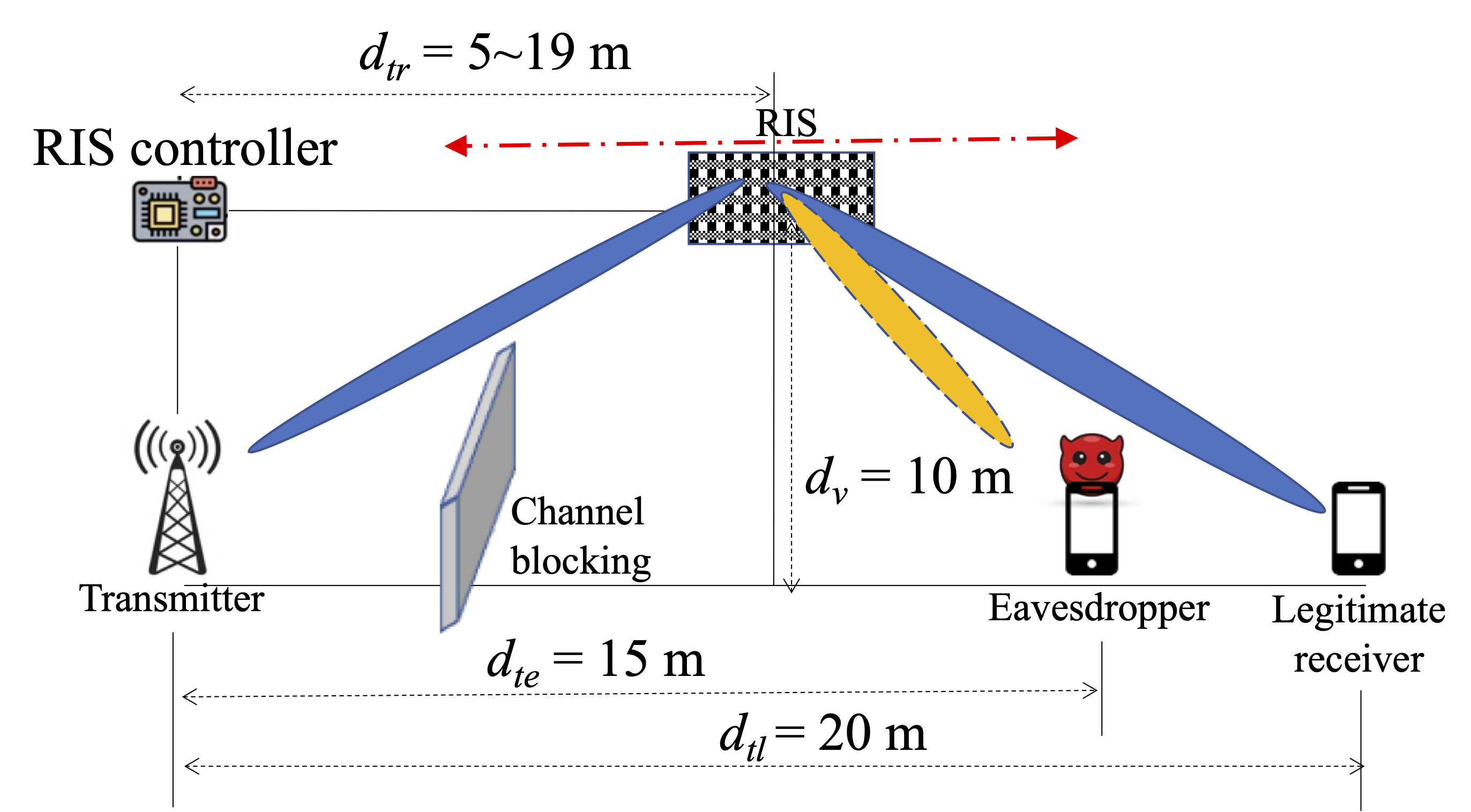}
\caption{Network topology for simulations.}\label{fig7}
\end{figure}

\section{Numerical Results}	
\label{secti6}

In this section, simulation results are presented to demonstrate the effectiveness of RISs in terms of PLS, particularly through the enhancement of the secrecy rate which is one of the critical and widely used measures of secure communication performance. {We examine the effects of the practical RIS model, the number of RIS elements, RIS beam design strategy, placement of RIS, quantized RIS phases, and AN generation on secrecy rate under the network topology depicted in Fig. \ref{fig7}.} Because the focus of simulations is to examine the secrecy rate improvement owing to RISs, a single-antenna scenario is considered as in \cite{Waqas_SENSOR20aa}. To facilitate communication between the transmitter and the receiver, an RIS with $N$ reflecting elements is deployed between them; its location may vary in the horizontal direction. Depending on the distance, the path loss model can be expressed as follows:
\begin{equation}
L(d)=C_o \left( \frac{d}{d_o} \right)^{-\gamma},
\end{equation}
where $C_o$ represents the path loss at the reference distance $d_o$, $d$ is the distance between the transmitter and the destination of a given link, and $\gamma$ is the path loss exponent. It is assumed that all the channel links follow a Rayleigh distribution \cite{Waqas_SENSrrd}. In the simulation setup, we set $C_o$ = -30 dB, $d_o$ = 1 m, and $\gamma$ = 3.0 or 3.5. There was assumed to be a noise variance of -100 dBm on the receiving end and a transmit power of 20 dBm on the transmitting end.

\begin{figure}
\centering
\includegraphics[width=1\linewidth]{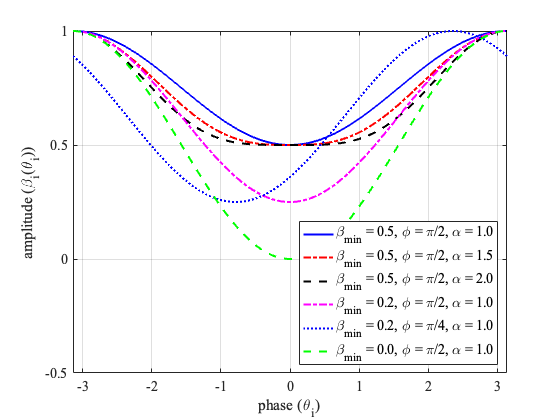}
\caption{Practical RIS model with different parameters: $\beta_{min}$, $\phi$, and $\alpha$.}\label{fig:amp_phase}
\end{figure}

{In an ideal RIS phase model, the amplitude is constant regardless of the phase, i.e., $\psi_i = e^{j\theta_i}$ for $i \in \{1, \cdots, N\}$. However, practical implementation and measurement results indicate that the amplitude of the RIS reflection coefficient varied with its phase. Specifically, the amplitude of the reflection coefficient of an RIS, $|\psi_i|  = \beta_i(\theta_i)$, is a function of $\theta$ as follows \cite{RIS_prctical:ICC20}.
\begin{align}\label{eq:amp_phase}
\psi_i & = \beta_i (\theta_i) e^{j\theta_i} \nonumber \\
          & =  \left[(1-\beta_{min})  \left( \frac{ \sin( \theta_i - \phi ) + 1}{2}\right)^\alpha + \beta_{min}\right] e^{j\theta_i},
\end{align} 
where $\beta_{min}$ is the amplitude level with maximal attenuation, that is, the minimum amplitude; $\phi$ is the phase difference between the phases at $\beta_{min}$ and $-\frac{\pi}{2}$; and $\alpha$ is the rate of amplitude attenuation. Fig. \ref{fig:amp_phase} illustrates several typical examples of phase-dependent amplitude responses. When $\beta_{min} = 1$ or $\alpha =0$, the amplitude becomes 1, regardless of $\theta_i$, i.e., $\beta_i(\theta_i) = 1$. Consequently, the phase-dependent amplitude model is transformed into an ideal RIS model. As the amplitude of an RIS is attenuated, that is, $\beta_i(\theta_i) \leq 1$, the quality of the RIS-reflected signal deteriorates.}

{The received signal at the receiver and the eavesdropper can be expressed by \cite{sdt4},
\begin{align}
y_l & = \sum_{i=1}^{N}h_i e^{j\theta_i} g_i s + z_l \\
y_e & = \sum_{i=1}^{N} h_i e^{j\theta_i} k_i s + z_e
\end{align}
where $h_i$ is a channel gain between the transmitter and the $i$th RIS element, $g_i$ and $k_i$ are channel coefficients from the $i$th RIS element to the receiver and the eavesdropper, respectively. $z_l$ and $z_e$ are additive white Gaussian noise with a variance $\sigma^2$ at the receiver and eavesdropper, respectively. As a measure of secrecy performance, the achievable secrecy rate can be expressed as follows:
\begin{align}
C_{s} = \max \left\{  \log_2 \left( 1+\frac{ \left\| \sum_{i=1}^N {h_i g_i e^{j\theta_i} }\right\|^2}{\sigma^2} \right)  \right. \nonumber \\
               \left. -\log_2 \left( 1+\frac{\left|\sum_{i=1}^N {h_i k_i e^{j\theta_i} }\right|^2}{\sigma^2}\right), 0 \right\}.
\end{align}
Assuming that the transmitter has the channel information from the transmitter to the receiver via RIS, i.e., $h_i$ and $g_i$ for all $i \in \{1, \cdots, N\}$, but no information on the eavesdropping link, the optimal phase to maximize the capacity of the legitimate link and the secrecy rate is given by
\begin{equation}
\theta_i = -\arg(h_i g_i).
\end{equation}
Consequently, all the RIS signals at the receiver have the same phase, and the received signal power is maximized. In this scenario, it is assumed that the eavesdropper is an out-of-network malicious device.
}

{
On the other hand, if the channel information on the eavesdropping link is available, the transmitter can take a pre-nulling policy to eliminate the transmit signal leakage to the eavesdropper and the whole information can be securely sent to the receiver. In this case, the transmitter is a base station (BS), and the receiver and eavesdropper are {user equipments (UEs)}. The BS sends downlink data to UEs in a time division multiple access (TDMA) manner. The receiver is the UE served in the current time slot while the eavesdropper was served in the previous time slot. The eavesdropper tries to overhear the UE in the current time slot, i.e., the receiver. It is called an ``in-network eavesdropper." When the channel information of the eavesdropping link is perfectly known at the transmitter, the RIS coefficient design for pre-nulling can be expressed as
\begin{align}\label{eq:null}
\boldsymbol{\xi}^T \boldsymbol{ \psi } = 0
\end{align} 
where $\boldsymbol{\xi} = [h_1k_1, h_2 k_2, \cdots, h_N k_N]^T$. If both amplitude and phase are controllable, the solution to \eqref{eq:null} is a vector in the null space of $\boldsymbol{\xi}^T$. When the elements of $\boldsymbol{\psi}$ have a unit amplitude, however, the solution to \eqref{eq:null} cannot obtained simply. An iterative algorithm in which an orthogonal projection and normalization processes are alternatively repeated can be used to obtain the solution to \eqref{eq:null} \cite{interference_nulling}. In practice, the channel information of the eavesdropping link may be outdated and then perfect pre-nulling of information leakage cannot be achieved. 
%For example, the correlation between the current eavesdropping channel gain and the previous channel information of the eavesdropping link at the BS can be expressed by 
%\begin{align}
%\boldsymbol{\xi}_{BS} = \mu \boldsymbol{\xi} + (1-\mu) \boldsymbol{\nu},
%\end{align} 
%where $\boldsymbol{\xi}_{BS}$ and $\boldsymbol{\xi}$ are the previous and current eavesdropping channel vectors, respectively, $\mu$ is the correlation coefficient and $\boldsymbol{\nu}$ is a complex Gaussian random vector with zero mean and unit covariance. 
}

As shown in Fig. \ref{fig7}, the transmitter, the eavesdropper, and the receiver are arranged in a line, and the RIS is located along a parallel line at a vertical distance of $d_v$ (= 10 m). The distance between the transmitter and the receiver is $d_{tl}$ (= 20 m), and the distance between the transmitter and the eavesdropper is $d_{te}$ (= 15 m). The horizontal distance between the transmitter and the RIS is $d_{tr}$, which can vary from 5 to 19 m. As the eavesdropper is closer to the transmitter than to the receiver, the eavesdropping link has a higher capacity than the legitimate link. Thus, a positive secrecy rate cannot be achieved without the RIS when the eavesdropping channel is not available at the transmitter. %The achievable secrecy rate is expressed as follows:

%As shown in Fig. \ref{fig7}, the transmitter, eavesdropper, and legitimate receiver are arranged in a line, and the RIS can locate along a parallel line at a vertical distance of $d_v$ (= 10 m). The distance between the transmitter and legitimate receiver is $d_{tl}$ (= 20 m) and the distance between the transmitter and eavesdropper is $d_{te}$ (= 15 m). The horizontal distance between the transmitter and RIS is $d_{tr}$ and it can vary from 5 to 19 m. 

%Since the eavesdropper is closer to the transmitter than the legitimate receiver, the eavesdropping link has a higher capacity than the legitimate link, thus the positive secrecy rate cannot be achieved without RIS. The achievable secrecy rate can be expressed as follows:
%\begin{equation}
%R_{sec}= \max \left(\log_2 \left( 1+\Gamma_r\right)-\log_2 \left( 1+\Gamma_e\right),0\right)
%\end{equation}
%where $\Gamma_r$ and $\Gamma_e$ represent the instantaneously received SINRs at the receiver and eavesdropper, respectively.

\begin{figure}
\centering
\includegraphics[width=1\linewidth]{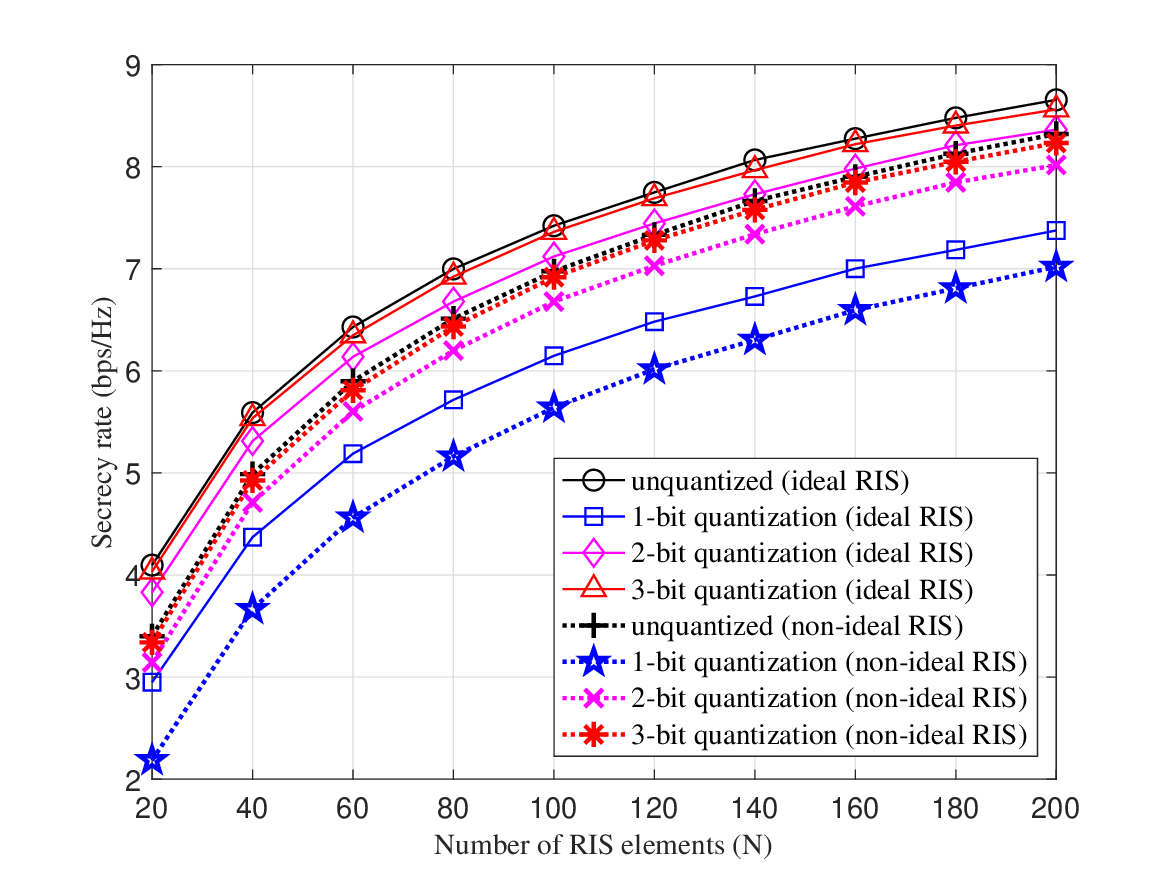}
(a)
\includegraphics[width=1\linewidth]{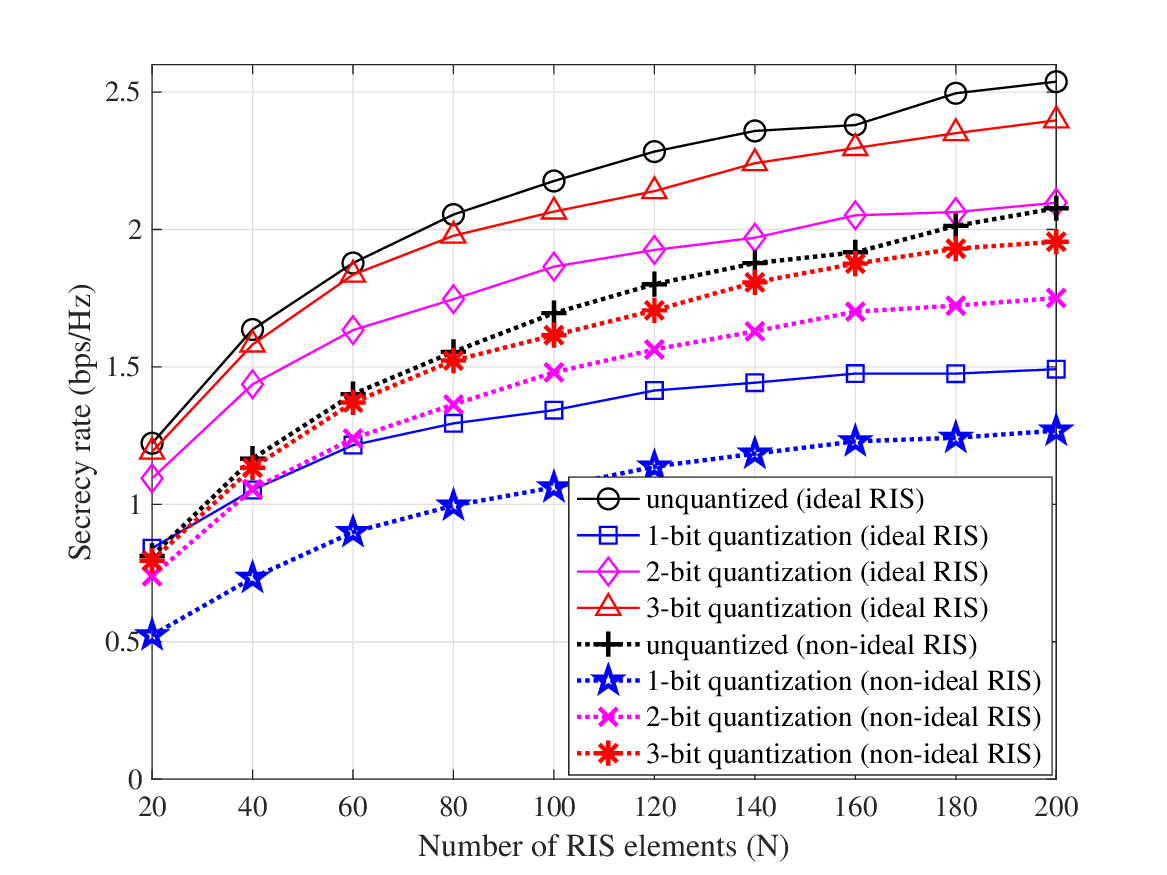}
(b)
\caption{{Secrecy rate versus the number of RIS elements ($N$) with different quantization levels of RIS reflection angles under ideal and non-ideal RIS models: (a) when the channel information of the eavesdropping link is unknown, (b) when the channel information of the eavesdropping link is known at the transmitter.}}\label{fig8}
\end{figure}

Figure \ref{fig8} illustrates the secrecy rate under ideal and non-ideal RIS models with $d_{tr}$ = 10 m and $\gamma$ = 3.0 {depending on the availability of eavesdropping channel information.} In the non-ideal RIS model, we set $\beta_{min} = 0.5$, $\phi = \pi/2$, and $\alpha$ = 2.0. Because of the obstruction located between the legitimate and eavesdropping links, the direct path from the transmitter is not considered when evaluating the secrecy rate.
%To control the RIS reflection pattern, the transmitter transmits the control information, for example, the reflection angle $\theta_i$ for $i = 1, \cdots, N$. Assuming that the channel coefficients between the transmitter and the $i$th RIS element, and the $i$th RIS element and the receiver are given by $h_i$ and $g_i$, respectively, the optimal phase to maximize the capacity of the legitimate link and the secrecy rate is defined as
%\begin{equation}
%\theta_i = -\arg(h_i g_i).
%\end{equation}
%Consequently, all the RIS signals have the same phase, and the received signal power is maximized. Because of the lack of channel information for the eavesdropping link, the signals from the RIS elements to the eavesdropper may have random phases.
To reduce the overhead of the control link between the transmitter and the RIS, the optimal phases are quantized as follows:
\begin{equation}
\hat{\theta_i} = Q_b(\theta_i) ,
\end{equation} 
where $Q_b(\cdot)$ denotes the quantization of $b$ bits. For example, $Q_1(\cdot)$ selected the phase closest to $\{\pi/2, -\pi/2\}$. 

Fig. \ref{fig8} shows that the secrecy rate increases with the number of RIS elements, and 3-bit quantization can accurately represent the performance achieved by the RIS under a given simulation environment. In addition, performance degradation is observed in the non-ideal RIS model. {When $N$ is large, the enhancement of the legitimate link quality with the RIS array gain is more effective in terms of secrecy rate than the pre-nulling of signal leakage in the eavesdropping link. Therefore, the secrecy rates in Fig. \ref{fig8} (a) are much higher than those in Fig. \ref{fig8} (b).} In this study, a uniform RIS codebook is assumed in both ideal and non-ideal RIS models. By designing the codebook and determining the RIS phases under a non-ideal RIS model, the loss of secrecy rate can be mitigated. However, this issue needs to be addressed in future studies.

\begin{figure}[!t]
\centering
\includegraphics[width=1\linewidth]{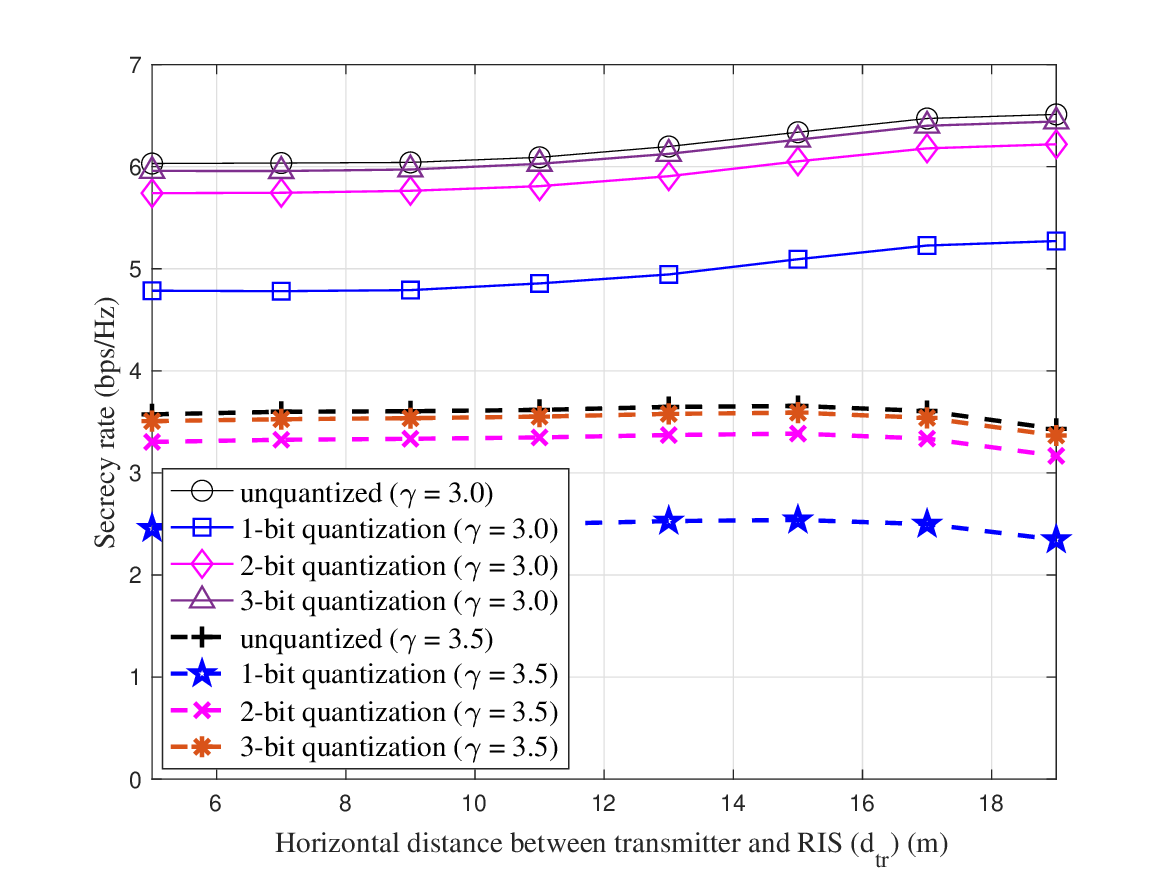}
(a)
\includegraphics[width=1\linewidth]{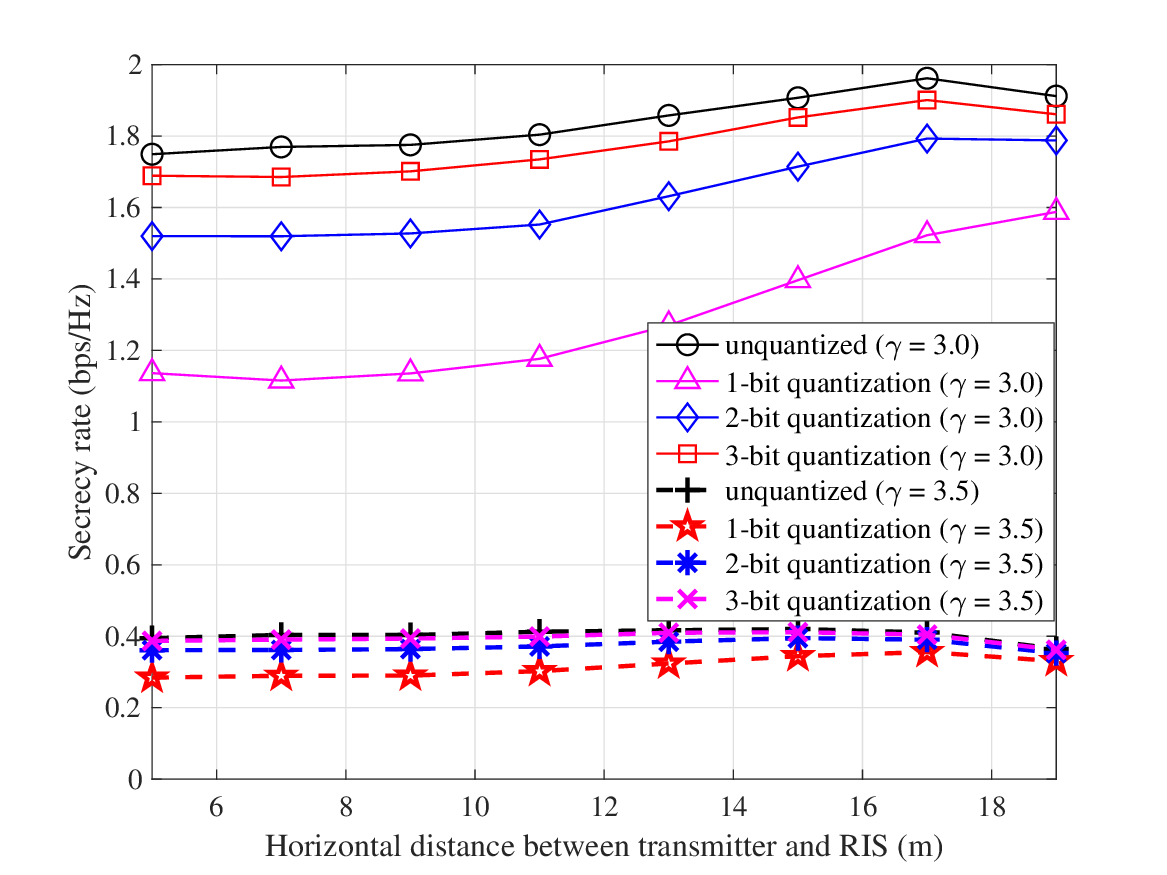}
(b)
\caption{{Secrecy rate versus RIS location, i.e., the horizontal distance between the transmitter and the RIS ($d_{tr}$), with different quantization levels of the RIS reflection angles and path loss exponents $\gamma \in \{3.0, 3.5\}$: (a) when the channel information of the eavesdropping link is unknown,  (b) when the channel information of the eavesdropping link is known at the transmitter.}}\label{fig9}
\end{figure}

{Fig. \ref{fig9}} shows the secrecy rate with respect to the RIS location, i.e., $d_{tr} \in [5, 19]$ m, {depending on the eavesdropping scenario} when the path loss exponents ($\gamma$) are $3.0$ and $3.5$. {Fig. \ref{fig9} (a) shows the secrecy rate with the out-of-network eavesdropper and Fig. \ref{fig9} (b) shows the secrecy rate with the eavesdropping channel information.} For this simulation, the number of RIS elements was 50, that is, $N$ = 50. When $\gamma$ = 3, the secrecy rate increases with $d_{tr}$. This is because the distance between the receiver and the RIS becomes shorter and the quality of the legitimate link becomes better than that of the eavesdropping link. In contrast, when $\gamma$ = 3.5, the secrecy rate decreases slightly with $d_{tr}$ because the effect of the link distance is more dominant than that of the RIS array gain. These results suggest that the path loss exponent has a greater impact on the secrecy performance than the location of the eavesdropper. Therefore, the RIS location can be optimized based on channel conditions, such as path loss exponents, to maximize the PLS performance. 

\begin{figure}[!t]
\centering
\includegraphics[width=1.0\linewidth]{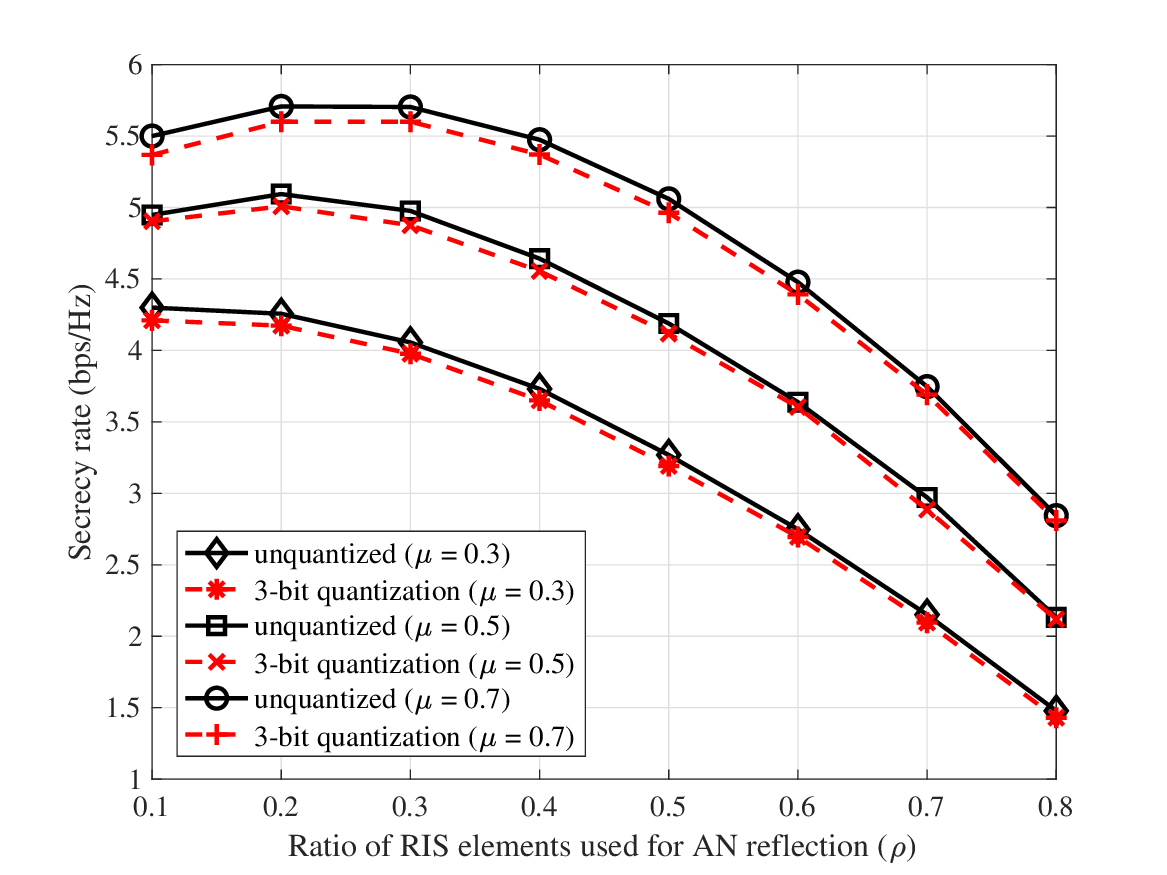}
\caption{{Secrecy rate versus the ratio of RIS elements used for AN reflection ($\rho$) with different power allocation factors to information signal $\mu \in \{0.3, 0.5, 0.7\}$ when $\gamma = 3.0$ and $N=100$.}}\label{fig10}
\end{figure}

{Fig. \ref{fig10} shows the secrecy rate achieved by utilizing AN and the corresponding RIS reflection pattern design strategy. To incorporate AN, it is assumed that one more antenna to transmit an AN signal is added at a transmitter. Therefore, an information signal and AN are separately transmitted by two transmit antennas with a power allocation ratio $\mu$. For example, when $\mu$ = 0.3 and the total transmit power is 100 mW (i.e., 20 dBm), the transmit power of an information signal is 30 mW and that of an AN signal is 70 mW. The RIS elements are partitioned into two groups with a ratio $\rho$: the RIS reflection angles of the first group with $\rho N$ RIS elements are determined to maximize the AN signal power to the eavesdropper while those of the second group with $(1-\rho) N$ RIS elements are determined to maximize the information signal power to the legitimate receiver, i.e., \cite{ww12}  
\begin{equation}
\theta_i = \left\{\begin{array}{ll} -\arg\{h_i k_i\}, & ~\mbox{for}~i \in \{1, \cdots, \rho N\}, \\
 -\arg\{h_i g_i\}, & ~\mbox{for}~i \in \{\rho N+1, \cdots, N\}. \end{array} \right.
\end{equation}
It is noteworthy that different RIS reflection design strategies for information and AN signals can be employed depending on the network condition. Fig. \ref{fig10} demonstrates the secrecy rate depending on the ratio of RIS elements used for AN reflection with the different power allocation factors for information and AN signals when $\gamma$ = 3.0 and $N$ = 100. It is shown that the optimal ratio of RIS elements for AN reflection ($\rho$) maximizing the secrecy rate increases when the power allocation ratio to information signal ($\mu$) increases. It means {that more} RIS elements should be used for AN reflection rather than information signal reflection when the higher power is allocated {to the} information signal to achieve the balance between information and AN signals. Such an optimal ratio under the unquantized RIS phase is also optimal in the quantized RIS phases.}

{In this section, we perform various simulations to investigate the impact of ideal/non-ideal RIS models, the number of RIS elements, RIS reflection pattern design depending on available CSI, RIS placement, unquantized/quantized RIS reflection phase, and AN power allocation/reflection on secrecy rate. Based on the provided results, we can conclude that RISs can provide an effective solution to PLS by optimizing various RIS design parameters in unfavorable channel conditions.}

\section{Conclusions}
\label{secti7}
In this study, we investigated RIS in the context of PLS, offering valuable insights into secure transmission design in the 6G era. First, we presented a comprehensive discussion of RIS and PLS. We examined RIS-enabled PLS designs, including beamforming, resource allocation, antenna/node selection, artificial noise, and cooperative communications. Furthermore, we clarified key research issues and their prospective solutions in the domain of RIS-enabled PLS, addressing channel estimation, beam configuration, resource management, strategic placement and passive information transfer for RIS, hardware and channel modeling, and optimization techniques. Additionally, we identified future research avenues highlighting ML-based solutions, advancements in RIS hardware, such as active RIS, and STAR-RIS, and security threats exploited by malicious RIS. Finally, numerical results were presented to demonstrate the effectiveness of RIS in improving PLS. Future research will focus on developing PLS solutions based on active RIS, STAR-RIS, and malicious RIS.

\section*{Acknowledgements}
This work was supported by Institute of Information \& communications Technology Planning \& Evaluation (IITP) grant funded by the Korea government(MSIT) (No.2020-0-00045, Development of movable high-capacity mobile communication infrastructure for telecommunication disaster and rescue).

%-\frac{2l^{4}\tan^{2}\alpha}{\cos\alpha})$ be $g(\alpha)$. \\
%Let $h=2l,2.5l,3l,3.5l,4l$ respectively, if there exists $\alpha_0$, $\alpha_1$ and $\alpha_2$ satisfying:\\
% \text{~~~~~~~~}$g(\alpha_1)'=0$, \\
% \text{~~~~~~~~}$g(\alpha_0)'<0$,\\
% \text{~~~~~~~~}and $g(\alpha_2)'>0$ \\
% where $\alpha_0$ is minor smaller than $\alpha_1$,$\alpha_2$ is minor larger than $\alpha_1$,
% then the expected $\alpha$ can be derived. Unfortunately, when  $g(\alpha_1)'=0$, $\alpha\notin\lbrack0,\frac{\pi}{2}\rbrack$.
% We substitute the series values of h to $g(\alpha)$, then
% achieve the minimal values of $g(\alpha)$ and their corresponding values of $\alpha$. The average value of $\alpha$ is 1.05.
% $\tan1.05=1.74$, so the depth of hole information announcement is:\\
% \text{~~~~~~~~}$1.74*l=1.74*\frac{L}{2}=0.87L$,\\
% Note  here $l$ is approximately represented by  $\frac{L}{2}$.
%
%%%%%%%%%%%%%%%%%
%%%%%%%%%%%%%%comment ends 12/25/14

%%%%%%%%%%%%%%%
%\cite{atakan2012body} is an example for a journal publication;\\
%\cite{russu2016secure}

\bibliographystyle{elsarticle-num}
\bibliography{egbib}

%\large{ Note: The manuscript was extended from the conference proceeding ``Jianjun Yang and Zongming Fei, HDAR: Hole Detection and Adaptive Geographic Routing for Ad Hoc Networks, Computer Communications and Networks (ICCCN), 2010 Proceedings of 19th International Conference on (pp. 1-6). IEEE, Zurich, Switzerland, August 2010."}

%% Authors are advised to use a BibTeX database file for their reference list.
%% The provided style file elsarticle-num.bst formats references in the required Procedia style

%% For references without a BibTeX database:

% \begin{thebibliography}{00}

%% \bibitem must have the following form:
%%   \bibitem{key}...
%%
 
% \bibitem{}

% \end{thebibliography}

\end{document}